\documentstyle[pre,aps]{revtex}

\newcommand{\psfig}[1]{{\small #1}}

\begin{document}
\title{Optical conductance fluctuations: diagrammatic
analysis in Landauer approach and non-universal effects} \author{M. C. W.
van Rossum$^a$, Th. M. Nieuwenhuizen$^a$, and R. Vlaming$^b$}
\address{$^a$Van der Waals-Zeeman Laboratorium, Universiteit van Amsterdam}
\address{$^b$Instituut Theoretische
Fysica, Universiteit van Amsterdam \\ Valckenierstraat 65, 1018 XE Amsterdam,
The Netherlands} \date{\today} \maketitle
\begin{abstract}
The optical conductance of a multiple scattering medium is
defined as the
total transmitted light of a normalized diffuse incoming beam. This
quantity, analogous
to the electronic conductance, exhibits universal conductance fluctuations.
We perform a detailed diagrammatic analysis of these fluctuations. With a
Kadanoff-Baym technique all the leading diagrams are systematically
generated. A cancellation of the short distance divergencies occurs, that
yields a well behaved theory. The analytical form of the fluctuations is
calculated and the theory is applied to optical systems. Absorption and
internal reflections, which are taken into account, reduce the fluctuations
significantly. \end{abstract}

\section{Introduction}\label{intro}
When studying the transmission of optical waves through random media one may
consider fluctuations in the {\em optical conductance}. Analogous to the
electronic conductance, this quantity is obtained when, first, the incoming
beam is monochromatic and diffuse, and second, all outgoing light is
collected. Since the electronic conductance is known to exhibit so called
universal conductance fluctuations (UCF), the same fluctuations are expected
in the optical conductance. Measurements of this quantity would constitute a
cornerstone in the analogy between optical and electronic mesoscopic systems.

We briefly review the situation for mesoscopic electron
systems\cite{umbach,altshuler2,lee,lee2,altshulerboekgeheel}. The electronic
conductance of mesoscopic samples is known to show reproducible sample to
sample fluctuations. Since the fluctuations are a consequence of scattering
from static impurities, they are static. Their magnitude is independent of
the sample parameters such as the mean free path, the sample thickness and
the average conductance. Hence they are called universal conductance
fluctuations. The mean conductance in the considered regime comes from
multiple scattered diffuse electrons. The UCF are a consequence of
interference of multiple scattered waves, causing correlations between two
diffuse paths. Therefore the fluctuations are much larger then one would
obtain classically by modeling the system by a random resistor array, in
which interference effects are neglected. That approach is valid only on a
length scale exceeding the phase coherence length, where the fluctuations
reduce to their classical value. The conductance fluctuates when the phases
of the waves in the dominant paths are changed. This happens, of course, if
one changes the position of the scatterers, eq. by taking another sample. One
may also keep the scatterers fixed but apply a magnetic field or vary the
Fermi energy. (In optical system one can vary the frequency of the light.) In
all these cases one modifies the phases of scattered waves, so that different
propagation paths become dominant.

Transport through mesoscopic systems is not only studied in electron systems,
but also using multiple scattered classical waves such as sound, microwaves,
and, particularly, light. The origin of the mesoscopic phenomena is the
interference of multiple scattered waves and in first approximation all
systems are described by the same equations, namely the scalar wave equation,
see for instance van Haeringen {\it et al.} \cite{vanhaeringen} for a review.
Thus one expects that similar large fluctuations in optical systems are
present in the diffuse transmission regime. An advantage of optical systems
over electronic ones is that optical systems are much cleaner: no equivalents
of phonons or electron-electron interactions are present. Indeed very
accurate measurements of the enhanced backscatter cone\cite{albada3},
correlation functions\cite{albada1,deboer} and intensity
distributions\cite{t3prl} were performed. Nevertheless, to the best of our
knowledge, the optical analog of the UCF has not yet been observed in optical
systems. Such experiments turn out to be difficult.  Although the magnitude
of the fluctuations is universal, they occur on a background of order $g$,
where $g$ is the dimensionless conductance (in optical experiments one
typically has $g\sim 10^3$). The relative value of the fluctuations to the
background is thus $1/g$, so that the $C_3$ correlation function is of order
$1/g^2$, typically of order $10^{-6}$. For electrons this problem is absent
as moderate values of $g$ are achievable. This is also the reason that
electrons are easier brought near Anderson localization, for which $g$ has to
take a critical value of order unity. In the electronic case the moderate
values of $g$, combined with very sensitive techniques for current
measurements, have led to many observations of the universal conductance
fluctuations. Recent optical experiments suggest, however, that the optical
analog of UCF should just be experimentally accessible, as the measurement of
the third cumulant in the total transmission was reported\cite{t3prl}. This
quantity is of the same order as the optical UCF, namely $1/g^2$. It is
expected that similar techniques can be applied to measure the optical UCF.
Microwave scattering is also interesting as it combines lower values of $g$
with many of the advantages of optical systems.

We are not only interested in just the size of the fluctuations, but also in
the somewhat more general frequency correlation. A change in the frequency
alters the interference pattern, just as occurs by changing the magnetic
field or the Fermi energy in the electronic case. It is known from
experiments that using the frequency as tunable parameter provides a good way
for measuring the fluctuations. In contrast to electronic systems, in optical
systems both angular resolved and angular integrated measurements of the
transmission are possible. Therefore various transmission quantities can be
measured in optical systems, each with its particular frequency correlation
function. In a recent review by Berkovits
and Feng\cite{berkovits3} the different correlations and their physical
interpretation are discussed extensively.
Denoting the transmission from incoming
channel $a$ (wave coming in under angles $\theta_a,\,\phi_a$)
to outgoing channel $b$ (waves transmitted into angles $\theta_b,\,\phi_b$)
as $T_{ab}$, the correlation
functions can be classified as \cite{berkovits3,feng}
\begin{equation}
\frac{\langle T_{ab}(\omega) T_{cd}(\omega+\Delta \omega) \rangle}
{ \langle T_{ab}(\omega)\rangle \langle T_{cd}(\omega+\Delta\omega)\rangle}
=1+C_1^{abcd}(\Delta \omega)+C_2^{abcd}(\Delta \omega)+C_3^{abcd}(\Delta
\omega)\label{eqdefc}. \end{equation}
The unity just comes from the product of averages. The $C_1$-term
in the correlation function is the most important one if the angular resolved
transmission $T_{ab}$ is measured. The reason is that $C_1$ is of order unity
when both the incoming directions $a$ and $c$, and the outgoing directions
$b$ and $d$ are, pairwise, close to each other. When dealing with one single
monochromatic plane wave ($a=c$),  $C_1$ describes the correlation of the
bright and dark speckle pattern. Diagrammatically, $C_1$ is the sum of all
reducible diagrams, that is to say, just as the unit contribution in
Eq.~(\ref{eqdefc}), it is equal to the product of two averages.

If instead of measuring light in one outgoing channel, all outgoing light is
integrated,  the sharply peaked and short ranged $C_1$ correlation function
is overwhelmed. By collecting the outgoing light, the total transmission
$T_a=\sum_b T_{ab}$ is measured; experimentally this is commonly done using
an integrating sphere\cite{deboer}. In this setup the $C_2$ correlation
function, which has a much smaller peak value but is long ranged, contributes
for all outgoing angles and becomes dominant. Its long range character arises
because, due to interference of the diffuse light paths, the outgoing
amplitudes are pairwise in phase. The $C_2$ correlation, which still depends
on the angles of the incoming beams $a$ and $c$, is of order $g^{-1}$. The
$C_2$ corresponds to a diagram where the two incoming diffusons interact
through a Hikami-vertex.

Finally, the $C_3$ term is dominant when the incoming beam is diffuse, and
all outgoing light is collected, so that, just as in electronic systems, the
conductance $g=T=\sum_{a,b} T_{ab}$ is measured. In that measurement
contributions where $a$ and $c$ are far apart are dominant. In contrast to
previous case, now also these incoming amplitudes must be pairwise in phase.
This occurs in a diagram where the two incoming diffusons interact twice, so
that a loop occurs; in principle, further loop insertions to it also
contribute. In spite of the fact that $C_3$ is of order $g^{-2}$, it is
dominates over the $C_1$ and $C_2$ terms as it has contributions for all
incoming and outgoing angles.

The $C_1$ and $C_2$ correlation have been studied in detail, both
experimentally\cite{deboer,garcia2,genack1} and
theoretically\cite{stephenboek}. It was also shown in experiments \cite{zhu}
and in theory\cite{lisyansky,skin} that absorption and internal reflection,
neglected in the earliest calculations, significantly reduce the
correlations. Among other methods, the $C_1$ and $C_2$ correlation were
successfully\cite{stephenboek,skin} calculated using a diagrammatic technique
based on the Landauer approach\cite{landauer,buttiker2}. One might hope that
the calculation of the $C_3$ or UCF in this approach is also straightforward.
It is well known, however, that the calculation in the Landauer approach is
quite cumbersome, since divergencies show up on scales of one mean free path
when the problem is treated on a macroscopic level using diffusons.

In order to circumvent these difficulties, one is tempted to use the Kubo
approach, often used in electronical systems to calculate the UCF
\cite{altshuler2,lee,lee2}. Furthermore, the results for the conductance
obtained by Kubo or Landauer formalism should be
identical\cite{fisher,janssen}. Yet the Kubo approach can not be applied
directly to optical systems, since it is not clear how external lines should
replace current vertices, and how absorption and internal reflections are to
be included. Therefore, we use the Landauer approach.

Technically, the difficulties in the Landauer approach are caused by the
vertices for partner exchange of two diffusons, the so-called Hikami
boxes.
Each Hikami box brings the square of the internal momentum, whereas the
current vertices are momentum-independent in the Kubo formula. As a result,
the
integral over the internal momentum of the closed loop is convergent in the
Kubo formula, while naively divergent in Landauer approach.

Two studies of the $C_3$ in the Landauer approach are known to us. In the
first, Kane, Serota and Lee\cite{kane} consider electronic systems and make
elegantly use of current conservation to derive an expression for the
correlation function. Although in optical systems the conserved quantity is
not the intensity but the energy, their prediction applies to optical systems
as well, since it amounts to a result for the same sums of scattering
diagrams, involving different parameters only. This result has not been
confirmed by a direct derivation, however. Moreover, since it relies on a
conservation law, it is not clear what happens when absorption is present.
The second study was performed by Berkovits and Feng\cite{berkovits3}. After
giving a very clear discussion of the problem, these authors calculated one
of the macroscopic diagrams (presented earlier by Feng, Kane, Lee and
Stone\cite{feng}) and subtracted the divergent parts by hand. In this way the
correct order of magnitude and the qualitative frequency dependence was
obtained.

It is our goal to clarify the situation by calculating the optical $C_3$
diagrammatically. A complete analysis of all leading diagrams is needed, in
which finally the divergent parts should cancel. We specialize to the case
where the conductance is measured. In this setup the amplitudes of the
incoming and outgoing diffusons are exactly in phase. If the $C_3$
correlation is measured as a (small) part of the correlation in the angular
transmission or total transmission, this phase condition need not be full
filled. Other contributions of order $1/g^2$ which are angular dependent are
then present. These contributions are both the diagrams presented below, with
different decay rates for the incoming and outgoing diffusons, but also new
diagrams contribute. In conductance measurements these complications do not
occur. Our fundamental approach immediately allows for inclusion of effects
due to boundary layers and absorption. Our calculations, although specialized
to optical systems, are valid for any mesoscopic system.

Tacitly we assumed that higher order correlations can be neglected. Altshuler
{\it et al.} \cite{altshulerboek} have shown that for small $n$ the $n$th
cumulants scale as $g^{2-n}$, so that for $g\gg 1$ the fluctuations have
indeed a Gaussian distribution; the far tail of the distribution is predicted
to be  log-normal. It would be interesting, however, to determine the full
distribution function of the fluctuations, as was done in experiment for the
angular resolved transmission\cite{garcia2} and the total
transmission\cite{t3prl}, and recently in theoretical work by Nieuwenhuizen
and Van Rossum\cite{probprl}.

The outline of this paper is as follows. First, we introduce the basics of
the diagrammatic technique and describe diffuse transport of light. In
section \ref{secset} we present the long distance diagrams and analyze the
divergencies arising from these diagrams in the diffusion approximation in
\ref{secdifdiv}. Next, in section \ref{secKB} we develop a Kadanoff-Baym
theory in order to generate all relevant scattering diagrams. In section
\ref{secshort} we show that the divergencies indeed cancel if all diagrams
are analyzed in detail. The general form of the fluctuations is calculated in
section \ref{seclong} and applied to optical systems, where absorption and
internal reflections may be present. The presentation is closed with a
discussion.

\section{Diffuse transport of light}\label{secdif}

In this section we discuss standard aspects of diffuse light transport, such
as the amplitude Greens function, the diffuse intensity and the transport
equation. We shall consider the situation of point scatterers in a medium
that has a dielectric constant different from its surroundings. We employ the
notation and results of the paper by Nieuwenhuizen and Luck\cite{thmn5}.
The main results of the present section are the expression
(\ref{C3-F}) for the $C_3$ correlation function, and the expression
(\ref{C3-G}) for the universal conductance fluctuations. These expressions
involve diagrams with incoming total-flux diffusons ${\cal L}_{\rm in}$
defined in (\ref{totfluxladin}), outgoing total-flux diffusons ${\cal
L}_{\rm out}$ defined in (\ref{totfluxladout}), and internal diffusons
${\cal L}_{int}$ defined in (\ref{Lint1}) or (\ref{Lint2}). Readers not
interested in the microscopic background of these relations may
skip the details of the present section.

We consider a quasi-one, quasi-two, or three
dimensional slab of thickness $L$ and area $A=W^{d-1}$, with ($W\gg L$).
which contains static, isotropic point scatterers.
As usual a scalar approximation is made for the electromagnetic field of the
light\cite{vanderhulst2}. For bulk properties this is justified
since the polarization is scrambled after a few scattering events.
For other applications, such as acoustic waves and spinless electrons,
the scalar property is immediate.
The scalar wave equation at given frequency $\omega$ reads
\begin{equation}
\nabla^2\psi(r)+\frac{\omega^2}{c^2} \epsilon(r)\psi(r)=0,
\end{equation}
where $c$ is the vacuum speed of light.
We shall consider a medium with dielectric constant $\epsilon_0$ and
density $n$ of small spheres with dielectric constant $\epsilon_2$
and radius $a_0$, located at random positions $R_i$. Going to the
limit of point scatterers we get
\begin{mathletters}
\begin{eqnarray} \epsilon(r)&=&\epsilon_0+\frac{4}{3}\pi a_0^3
(\epsilon_2-\epsilon_0)\sum_i\delta(r-R_i)\qquad 0<z<L, \\
&=&\epsilon_1 \qquad \qquad \qquad \qquad\qquad
\;\;\;\;\;\;\;\;\;\;\;z<0;\; z>L, \end{eqnarray}
\end{mathletters}
where $\epsilon_1$ is the dielectric constant  in the surrounding medium.
The wave numbers in the surrounding medium and in
the random medium are, \begin{equation}k_1=\frac{\omega}{c}
\sqrt{\epsilon_1}
, \qquad k=mk_1\equiv \sqrt{\frac{\epsilon_0}{\epsilon_1}}k_1,
\end{equation} respectively.
In the bulk the average retarded Greens function (amplitude Greens function)
in three dimensions with momentum ${\bf p}$ reads
\begin{equation}\label{greenfie}
G({\bf p})=\left( {\bf p}^2 -k^2 -nt \right)^{-1}. \end{equation} For small
scatterers the scattering becomes isotropic with an
effective strength described
by the $t$-matrix\cite{thmn4} \begin{equation}
t=\left(\frac{1}{u}-\frac{1}{u_0}-
i\frac{k}{4\pi}\right)^{-1}, \label{eqdeft} \end{equation} where $u=4 \pi
k^2 a_0^3
(\epsilon_2-\epsilon_0)/3$ is the bare scattering strength and $u_0$ is an
internal parameter of the point scatterer. The mean free path is defined as
\begin{equation}\ell =\frac{4\pi}{nt \overline{t}}\, ,\end{equation} in which
$n$ is the scatterer density. The metallic mesoscopic regime is defined by: $
k\ell \gg 1$ and $\ell \ll L\, $. Eq.(\ref{eqdeft}) satisfies the optical
theorem ${\rm Im}(t)=k\bar tt/4\pi a$, where $a$ is the albedo of the
scatterer. The case of no absorption, which we treat first, corresponds to
$a=1$. In electronic systems absorption does not take place, but a similar
behavior arises from de-phasing; in that case the phase coherence length
plays the role of absorption length and, for our purpose, has to be larger
than the system size.

For disordered electron systems one deals with weak $s-$wave scattering and
the diagrammatic expansion can be carried out in second order Born
approximation. In experiments on optical systems however, efficient
scattering is achieved by taking resonant scatterers and strictly one now has
to calculate the full Born series\cite{thmn4}. In the main part of this work
we study how the cancellation of some short range contributions takes place.
This we do in the second order Born approximation, where $t\approx
u+iu^2k/4\pi$ and $\ell\approx 4\pi/nu^2$. Working with the full Born series
would introduce a large number of extra classes of diagrams so that the total
number of diagrams increases dramatically. We expect, however, that the
cancellation of divergencies, when shown in the second order Born
approximation, also holds if the full Born series is considered. It is widely
expected and confirmed in one of our earlier papers that for static
quantities
the difference between the two should only be a renormalization of the mean
free path in the final results\cite{hik}. We thus continue to work with the
full Born series where possible. The cancellation of short range divergencies
will only be shown within the second order Born approximation.

\begin{figure} \centerline{\psfig{file=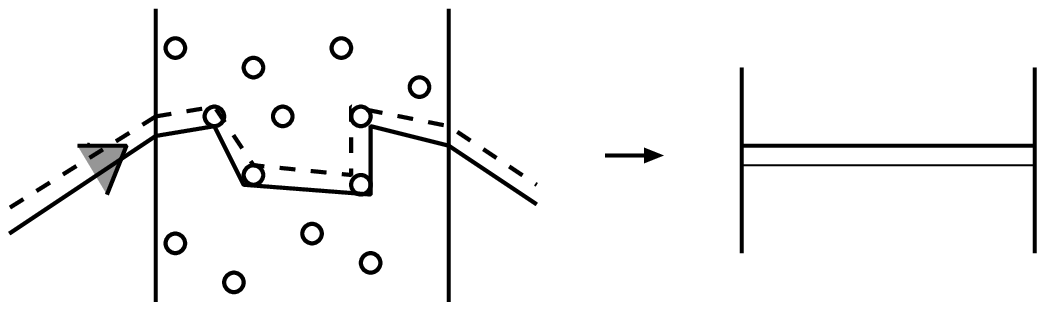,width=8cm}}
\caption{Left: an example of an actual scattering process; a retarded (full
line) and an advanced amplitude (dashed line) come from the left and share
the same path through the sample. Right: schematic representation of the
average process, the diffuson.} \label{figt1} \end{figure}

Intensity transport in transmission is dominated by ladder diagrams or
diffusons. A diffuson is made up by pairing one retarded and one advanced
propagator sharing the same path through the sample, as is depicted in
Fig.~\ref{figt1}.

Although the transport deep in the bulk is accurately described with the
diffusion equation, the precise behavior near
the (reflecting) surface has to be derived from the Schwarzschild-Milne
integral-equation\cite{vanderhulst2}.
Consider a plane wave with unit
flux impinging on the sample under angle $\theta_a$
\begin{equation} \psi_a^{\rm in}(r)=
\frac{1}{\sqrt{A k_1\cos\theta_a}}\;
{\rm e}^{i {\bf Q_a}\rho+ik_1\cos\theta_a z}, \;
 z<0 ,\end{equation}
 where $\rho=(x,y)$ is the transversal coordinate and
$Q_a=k_1\sin\theta_a (\cos\phi_a,\sin\phi_a)$
is the two dimensional transverse momentum of the incoming beam.
Inside the slab the unscattered part of the intensity decays over
one mean free path. This is described by (\ref{greenfie}) since its
pole lies slightly away from the real axis, due to the imaginary part
of $t$. The source of diffuse intensity is the first scattered
intensity.
\begin{equation}\label{Sa}
S_a(z)=T(\mu_a) \; nt\overline{t} \; |\psi_a^{\rm in}(r)|^2
=\frac{4\pi T(\mu_a)}{A k_1\mu_a\ell} \,{\rm e}^{-z/\ell \mu_a},
\end{equation}
where $\mu_a=\cos\theta_a'$ involves the angle $\theta_a'$ of the refracted
 beam with respect to the $z$-axis. The source is proportional to
 the intensity transmission coefficient of the boundary
between the two dielectrics
\begin{equation} T(\mu)=\frac{4\mu\sqrt{\mu^2-1+1/m^2}}
{[\mu+\sqrt{\mu^2-1+1/m^2}]^2}.\label{eqdefTmu}\end{equation}
The twice scatterered intensity follows
from $S_a$ as $nt\bar t \int dr' |G(r,r')|^2 S_a(r')$. The geometric sum
of first, second, third,..., times scattered intensity
is the multiple scattered or diffuse intensity ${\cal L}_a$. It is
generated from the first scattered intensity by the Schwarzschild-Milne
equation \begin{equation} {\cal L}_a (z)=S_a(z)+\int_0^L dz' M(z,z')
{\cal L}_a(z'). \label{eqmilne} \end{equation}
The kernel $M$ reads
\begin{equation}
M(z,z')=M_B(z-z')+M_L(z+z')+M_L(2L-z-z'), \end{equation}
which contains the bulk term
\begin{equation} M_B(z-z')=\int_0^1 \frac{d\mu}{2\mu} {\rm e}^{-|z-z'|/\mu
\ell} , \end{equation}
and layer terms describing internal reflections at the interfaces at
$z=0$ and $z=L$. The layer term reads \begin{equation} M_L(z+z')=\int_0^1
\frac{d\mu}{2\mu}(1-T(\mu)) {\rm e}^{-(z+z')/\mu \ell}. \end{equation} It
involves the total path length of the radiation going from depth $z$ to the
boundary at $z=0$, reflected there and going to depth $z'$. In the bulk of
the slab (several mean free paths away from the boundaries) the diffuse
intensity has a simple behavior\cite{thmn5} \begin{equation} {\cal L}_a^{\rm
in}(z)=\frac{4\pi T(\mu_a)\tau_1(\mu_a)} {k_1\ell m\, A\mu_a}
\,\frac{L+z_0-z}{L+2z_0} .\label{Lain} \end{equation} Here $\tau_1$ is a
factor that determines the limit intensity of a semi-infinite slab; of
course, it depends on the incident angle. The length $z_0$ is called the
``injection depth''. For isotropic scattering one has\cite{vanderhulst2}
$z_0=0.7104 \ell$. However, it becomes larger when there is a mismatch of the
indices of refraction between the scattering medium and the
surroundings\cite{thmn4}. Note that (\ref{Lain}) satisfies the diffusion
equation $\nabla^2{\cal L}_a^{\rm in}=0$; the more complicated
Schwarzschild-Milne equation is needed to fix the parameters of its solution.

In the electronic conductance measurements, however, waves coming from all
directions are involved. In the Landauer formula for the conductance,
\begin{equation}
G=\frac{2e^2}{h}\sum_{ab}T_{ab},\end{equation}
one needs to sum the
transmission coefficients $T_{ab}$ of  waves with unit flux coming in
channel $a$ and going out to channel $b$.
In an optical experiment, however, the
incoming diffuse beam may have angular weights that differ by a factor of
order unity. When using integrating spheres one measures the outgoing
intensity, rather than the outgoing flux. As compared to the electronic
case, it does not bring a factor $\mu_b$ to the weight of the outgoing
channel. As these differences between electronic and optical measurements
only lead to different numerical pre-factors, we calculate the optical UCF
with the same weights as in the electronic case. Summing (\ref{Sa})
over the channels $a$  yields a source for the diffuse intensity
\begin{equation} S(z)=\sum_aS_a=\frac{2k}{\ell}
\int_0^1 d\mu {\rm e}^{-z/\ell \mu_a} .\end{equation}
This is again the input in the
Schwarzschild-Milne equation(\ref{eqmilne}).
In the bulk the intensity now has the diffusive behavior
\begin{equation}\label{totfluxladin}
{\cal L}^{\rm in}(z)=\frac{ 4k}{\ell}\,\frac{L+z_0-z}{ L+2z_0}.
\end{equation}
 This object has been termed the incoming
{\it total-flux diffuson}\cite{probprl}. It has the same depth dependence
as ${\cal L}_{\rm in}^a$, but contains a different prefactor. Defining
\begin{equation}\epsilon_a=\frac{\pi T(\mu_a)\tau_1(\mu_a)}{k_1^2 m^2
A\mu_a}, \end{equation} we can verify the sum rule \begin{equation}
\sum_a\epsilon_a=\int \frac{d^2 Q}{(2\pi^2)} \frac{\pi
T(\mu)\tau_1(\mu)}{k_1^2m^2\mu}= \frac{1}{2}\int d\mu
T(\mu)\tau_1(\mu)=1.\end{equation} Here we used that $k=mk_1$, the definition
$\mu=\sqrt{1-Q^2/k^2}$ and (2.30) of Ref.\cite{thmn5}. One thus has
\begin{equation} \label{tfaa} {\cal L}_{\rm in}^a(z)=\epsilon_a {\cal L}_{\rm
in}(z). \end{equation}

On the outgoing side, radiation emitted at a point $r=(\rho,z)$ inside
the slab will
propagate to a point $(Z,\rho')$ outside the sample ($Z>L)$
as described by the Greens function of semi-infinite medium with dielectric
function $\epsilon({\bf r})=\epsilon_0$ for $z<L$ and $\epsilon({\bf
r})=\epsilon_1$ for $z>L$ \begin{equation} G(\rho,z;\rho',Z)=
\frac{1}{A}\sum_Q G(z;Z;Q){\rm e}^{iQ(\rho-\rho')} ,\end{equation}
in which  $G(z;Z;Q)$ is the 1-d Fourier transform of
Eq.~(\ref{greenfie}),
\begin{equation} G(z;Z;Q)=
\frac{i}{P+p} {\rm e}^{iP(L-z)+ip(Z-L)}\,, \qquad P=\sqrt{k^2-Q^2+nt},\qquad
p=\sqrt{k_1^2-Q^2}. \end{equation}
In the far field ($Z\gg L$) the total transmitted intensity reads
\begin{equation}
\int d^2\rho'|G(\rho,z;\rho',Z)|^2=\frac{1}{A}\sum_Q|G(z;Z;Q)|^2
\end{equation}
Since $p_z=k_1 \cos\theta_a =k\mu_a$, the according flux is
\begin{equation}
\Phi(z)=\frac{1}{A}\sum_Q k\mu |G(z;Z;Q)|^2=\frac{k}{8\pi}\int_0^1
d\mu {\rm e}^{-(L-z)/\ell\mu}\, .\end{equation}
It leads to a source $S(z)=4\pi\Phi(z)/\ell$ in the Schwarzschild-Milne
 equation.  The outgoing total-flux diffuson in transmission therefore reads
\begin{equation}\label{totfluxladout}
{\cal L}_{\rm out}(z)=
\frac{k}{\ell}\,\frac{z+z_0}{ L+2z_0}. \end{equation}
Apart from a reflection, this expression differs by a factor 4
from (\ref{totfluxladin}). The pre-factors would be
the same if our Greens functions were multiplied by a factor 2; this
amounts to the same as taking a kinetic term $\nabla^2/2$ rather than
$\nabla^2$, such as occurs
in electronics in units where $\hbar=m=1$.

For outgoing intensity in direction $b$ one has, in analogy with
(\ref{tfaa}), \begin{equation} \label{tfbb} {\cal L}_{\rm
out}^b(z)=\epsilon_b {\cal L}_{\rm out}(z) \end{equation} We call the
incoming and outgoing diffusons, {\em external} diffusons, because they are
connected to the outside of sample. This in contrast to the {\em internal}
diffusons that begin and end at interference vertices inside the medium. Away
from the surface the internal diffusons obey the well known diffusion
equation: $\nabla^2 {\cal L}_{int}(r)=12\pi\delta(r-r') /\ell^3$.

The result for the angle-resolved transmission of
Nieuwenhuizen and Luck \cite{thmn5} can be written as
\begin{equation} \langle T\rangle_{ab}=\epsilon_a\epsilon_b \langle
T\rangle, \end{equation} where $ \langle T\rangle$ is the average
conductance \begin{equation}\label{gtt}
\langle T\rangle =\frac{k^2 A\ell}{3 \pi(L+2z_0)}.\end{equation}
in dimensionless units. Restoring units for the electronic case one has for
the average conductance \begin{equation} \langle G\rangle
=\frac{2e^2}{h}\,\frac{k^2 A\ell}{3 \pi(L+2z_0)}.\end{equation}

All the above can be generalized to include absorbing scatterers,
frequency differences between incoming beams.
For an electronic system the correspondent effects
would be a change in Fermi energy, rather than a change in frequency,
and the effect of a finite incoherence length, rather than the
effect of absorption. For this more general case the diffusion
equation reads
\begin{equation}[-\nabla^2+ \kappa^2+i\Omega ]{\cal L}_{int}(r)
=\frac{12\pi}{ \ell^3}\delta(r-r').\end{equation}
The inverse absorption length $\kappa$ is related to the
albedo $a$, as $\kappa^2=3(1-a)/ \ell^2$; $\Omega=\Delta \omega/D$ is the
ratio
of the frequency difference of the amplitude propagators and the diffusion
constant $D$. It holds that $D=\frac{1}{3} v_E \ell$, where $v_E$ is
transport speed\cite{albada2}. In the slab
geometry after Fourier-transform in the $z$-direction the diffusons obey:
\begin{equation} -\frac{d^2} {d z^2} {\cal
L}_{int}(z,z';M)+M^2 {\cal L}_{int}(z,z';M)=\frac{12 \pi}{\ell^3} \delta(z-z')
\label{difeq},\end{equation}
where we have defined the decay rate
\begin{equation}M^2=Q^2+\kappa^2+i\Omega. \label{eqdefm} \end{equation}
The complex parameter $M$ describes the exponential decay ${\cal L}\sim
\exp(-Mz)$ of the diffuse intensity
in the $z$-direction. The solutions to the
diffusion equation are a linear combination of hyperbolic sines and
cosines. In conductance measurements using integrating spheres
external diffusons with momentum or frequency terms yield no contribution,
and thus $M=\kappa$ for the external diffusons, i.e.
 \begin{mathletters}
 \begin{eqnarray}
 {\cal L}_{\rm in}(z)&=&\frac{4k }{\ell}
\frac{ \sinh\kappa(L-z)+ \kappa z_0 \cosh \kappa(L-z) }{(1+\kappa^2
z_0^2)\sinh \kappa L+2\kappa z_0 \cosh \kappa L }, \\
 {\cal L}_{\rm out}(z)&=&\frac{k }{\ell}
\frac{\sinh  \kappa z+\kappa z_0 \cosh \kappa z}{(1+\kappa^2
z_0^2)\sinh \kappa L+2\kappa z_0 \cosh \kappa L }.
\label{eqliokappa}
\end{eqnarray}\end{mathletters} Whereas for the internal diffusons
 \begin{equation}\label{Lint1} {\cal L}_{int}(z,z';M)=\frac{12 \pi}
{\ell^3}\frac{[\sinh M z+M z_0 \cosh M z ]
[ \sinh M(L-z')+ Mz_0 \cosh M(L-z') ]}{(M+M^3 z_0^2)\sinh M L+2M^2
z_0 \cosh ML } ,\end{equation} where it is assumed that $z<z'$, otherwise
$z$ and $z'$ must be interchanged on the r.h.s.
With equal indices of refraction inside and outside the sample the
extrapolation length $z_0$ is, as stated above, $0.7104 \ell$ and
 thus the terms
involving $z_0$ yield
contributions of the order $\ell/L$. For optically thick samples ($L\gg
\ell$) one has \begin{equation}\label{Lint2}
 {\cal L}_{int}(z,z';M)=\frac{12 \pi}{\ell^3}
\frac{\sinh M z\sinh M(L-z')}{M\sinh M L} ,\end{equation}
Consistent with this expression the boundary conditions become approximately
 ${\cal L}_{int}(0,z')={\cal L}_{int}(L,z')=0$.
If there is an index mismatch, the surfaces partially reflect and  the
extrapolation length $z_0$ increases\cite{thmn5}.
Internal reflection becomes especially important when
$z_0$ becomes comparable to the sample thickness. Such may occur for
large index mismatch and moderate thicknesses.

The $C_3$ correlation function, defined in (\ref{eqdefc}), involves incoming
diffusons ${\cal L}_{\rm in}^{a,c}$, and outgoing ones ${\cal L}_{\rm
out}^{b,d}$.
Due to the factorization of external
direction dependence, see (\ref{tfaa}), (\ref{tfbb}), and
(\ref{gtt}), it cancels from $C_3$. We can write
\begin{equation}\label{C3-F}
C_3^{abcd}(\kappa,\Omega)=C_3(\kappa,\Omega)
=\frac{1}{\langle T\rangle^2}\sum_Q F(Q,\kappa,\Omega)  .
\end{equation}
where $Q$ is the two-dimensional transversal momentum. The function $F$
is the main object to be determined in this paper. It
is thus calculated at fixed $Q$ and with external diffusons being
total-flux diffusons. One finds from
(\ref{eqdefc}) and (\ref{C3-F}) the conductance fluctuations
\begin{mathletters}
\begin{eqnarray}
C_T(\kappa,\Omega)&=& \langle T(\omega) T(\omega+\Delta \omega) \rangle
-\langle T(\omega)\rangle \langle T(\omega+\Delta \omega) \rangle \nonumber
= \sum_{abcd} \langle T\rangle_{ab}\langle T\rangle_{cd}C_3 \nonumber \\
&=& \sum_Q F(Q,\kappa,\Omega) \nonumber \\
&=& F(0,\kappa,\Omega) \qquad \qquad \mbox{quasi 1-d}, \\
&=& W \int \frac{{\rm d}
Q}{2\pi} F(Q,\kappa,\Omega)\qquad\mbox{quasi 2-d},
\\ &=& W^2 \int \frac{{\rm d}^2  Q}{(2\pi)^2}
F(Q,\kappa,\Omega)\qquad\mbox{3-d}. \label{eqt2} \end{eqnarray}
\end{mathletters}
For electronic systems one finds \begin{equation} \label{C3-G} \langle G(k)
G(k+\frac{1}{3}\ell \Omega) \rangle-\langle G(k) \rangle\langle
G(k+\frac{1}{3}\ell \Omega) \rangle=\left(\frac{2e^2}{h}\right)^2
C_T(0,\Omega) .\end{equation} These results can be extended for other
geometries. If the width is comparable to the thickness of the slab, the
momentum integral discretizes into a sum over transversal
eigenmodes\cite{lee2}. The result can be generalized further  to arbitrary
geometries by taking $x,y$ dependence into account, and calculating the
diffusons using appropriate boundary conditions.

\section{Long range diagrams}\label{secset}
At this point we address the structure of the leading diagrams for the
correlation functions, defined in (\ref{eqdefc}). For all diagrams there are
two incoming advanced fields, which we momentarily term $i$ and $j$, and two
retarded ones, $i^\ast$ and $j^\ast$. The first term on the r.h.s. in
Eq.~(\ref{eqdefc}) follows from the diagram where $i$ and $i^\ast$ are paired
into an incoming diffuson, and the same for $j$ and $j^\ast$. These diffusons
have no common scatterers, so for this contribution the expression
factorizes into a product of averages. For the $C_1$ term in (\ref{eqdefc})
such a factorization also takes place\cite{feng}. However, in this term the
pairings are $ij^\ast$ and $ji^\ast$. In the $C_2$ correlation function there
are two terms. In the first the incoming diffusons have pairings $ij^\ast$
and $ji^\ast$. These diffusons interfere in some point in space, where they
exchange partners. The outgoing pairings are then $ii^\ast$, $jj^\ast$. It is
this term that contributes in measurements of the total transmission. The
time-inversed diagram, with in- and outgoing part of the diagram
interchanged, also contributes to Eq.~(\ref{eqdefc}). However, it does not
contribute when this expression is summed over $b$ and $d$, i.e. in total
transmission.

Interference of diffusons occurs if they exchange an amplitude in the
presence of a common scatterer. In a diagrammatic language this is described
by so-called Hikami-vertices\cite{gorkov,hikami}. They are depicted as shaded
polygons, the four point vertex represents the interference of four
diffusons, the six point vertex connects six diffusons. The shaded polygons
in the figure indicate that the vertices are dressed, in second order Born
approximation the dressed four point vertex is the sum of three diagrams, the
six point vertex is the sum of sixteen diagrams\cite{hikami}.

In electronic experiments where one measures the conductivity, and in optical
experiments where one uses an integrating sphere on the incoming side for
creating a diffuse beam, the two amplitudes in the diffuson must have exactly
the same phase.  Therefore, the incoming diffusons cannot have a momentum or
frequency difference, and the pairing must be $ii^\ast$ and $jj^\ast$.
In Fig.~\ref{figreg}(a) the incoming diffusons interfere somewhere in the
slab.
In  a diagrammatic language the diffusons interchange a propagator so that
the pairing is changed into $ij^\ast$ and $ji^\ast$. Propagation continues
with these diffusons, which, due to the different pairing can have non-zero
frequency difference and non-zero momentum.  But in order to be dominant, the
outgoing diffusons can also not have a momentum or frequency difference.
Therefore, somewhere else in the slab a second interference occurs.
Again exchanging an amplitude, the original pairing, $ii^\ast$ and $jj^\ast$,
is restored and the two diffusons propagate out, see Fig.~\ref{figreg}(a)i.
Some other contributions occur as well. Whereas the incoming and outgoing
pairings are always $ii^\ast$ and $jj^\ast$, the internal ones may be
different. In Fig.~\ref{figreg}(a)ii the first incoming diffuson meets an
outgoing diffuson and amplitudes are exchanged. These
internal diffusion lines meet at a second point where the original pairings
are restored. It is clear that in this process the intermediate paths are
traversed in time reversed order. Due to time-reversal symmetry they give a
similar contribution as previous Fig.~\ref{figreg}(a)i. In
Fig.~\ref{figreg}(b)i, for instance, a diffuson breaks up such that one of
its amplitudes makes a large detour, returns to the breaking point and
recombines into an outgoing diffuson. The second incoming diffuson crosses
this excess path of the amplitude, and one of its amplitudes follows exactly
the same contour as the one of the first diffuson. The fourth amplitude
resides and finally recombines with its original partner amplitude to form an
outgoing diffuson. Finally, Fig.~\ref{figreg}(c) depicts the situation where
only one internal diffuson occurs. Its endpoints must lie within a distance
of a few mean free paths. Because of its local character this class does not
show up in the final result; it is needed, however, in the regularization
process, since it contains terms that cancel divergencies from the other two
classes.

Due to the diffusive behavior of the internal propagators we call all these
the ``long range diagrams''. Their internal lines are diffusons (ladder
diagrams), that is to say, in these lines there can be an arbitrary number of
scatterers. Note that these long range diagrams also include terms with only
a small number  of scatterers, e.g. one or two. The latter contributions are
of course not really long range; however they contribute to the geometric
series that represents the ladder diagram. We state this explicitly, since
below we will discuss some unexpected problems of these short ranged
contributions to the long range diagrams. There are also some special short
ranged contributions. Due to various subtleties, they resist a general
treatment; their calculation is postponed to section \ref{secshort}.

Another class of diagrams can be constructed by taking the upper half of
Fig.~\ref{figreg}(a)ii and combining it with two lower half of
Fig.~\ref{figreg}(b)i. This diagram contains again two four-point vertices
and also two internal diffusons. It is easily seen that incoming amplitudes
$ii\ast$ and $jj\ast$ are mixed into  $ij\ast$ and $ji\ast$. These new
diagrams are $1/g^2$-corrections to the angular dependent $C_2$ and as such
do not contribute in conductance measurements.

\section{Divergencies in the diffusion approximation}\label{secdifdiv}

\begin{figure} \psfig{file=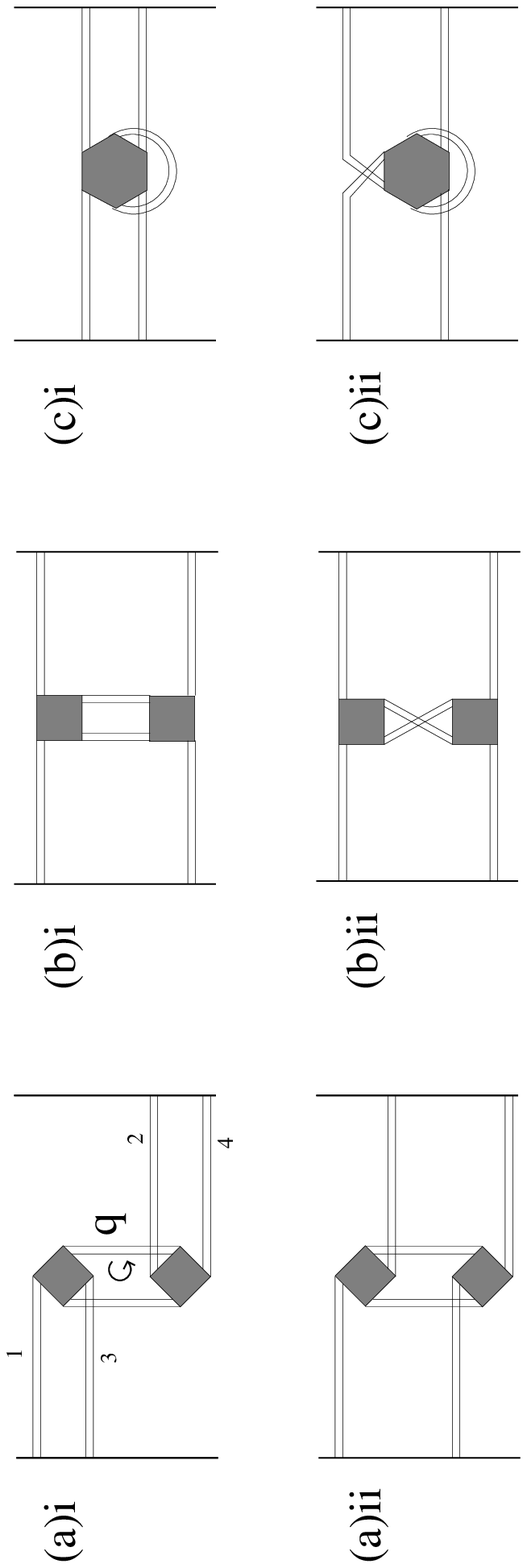,width=16cm}
\caption{ The leading contributions to the conductance fluctuations, apart
from some special short distance processes dealt with in section
\protect{\ref{secshort}}. The incoming diffusons from the left interfere
twice before they go out on the right. The close parallel lines correspond to
diffusons; the shaded boxes are Hikami vertices; ${\bf q}$ denotes the free
momentum which is to be integrated over.} \label{figreg} \end{figure}

The diagrams for the conductance fluctuations contain a loop; the two
internal diffusons have a free momentum, over which one has to integrate. In
Fig.~\ref{figreg}(a) this momentum is denoted ${\bf q}$. Physically, one
expects
important contributions to the conductance fluctuations if the distance
between the two interferences vertices ranges from the mean free path to the
sample size. But in this section we will show that the ${\bf q}-$integral
for the long range diagrams diverges for large momentum, i.e. when the two
interference processes are close to each other. The standard picture of
diffuse transport with diffusons and interference described by Hikami
vertices, that works so well for loop-less diagrams such as the $C_2$
correlation function\cite{stephenboek}, and the third cumulant of the total
transmission\cite{t3prb}, now becomes spoiled by these divergencies. In
section\ref{secshort} we solve the problem by going back to mesoscopic
scales, and consider all scattering events.

The problem becomes clear if we calculate the diagrams of Fig.~\ref{figreg}.
First, we need the expressions for the Hikami vertices. In order to derive
the vertices, the momenta are expanded to leading order in $({\bf q}
\ell)$. For
the large ${\bf q}\ell$ this is in principle not allowed, but in practice it
could still work.
The four point vertex and the six point vertex are found by summation of the
bare vertex and its dressings. The calculation has been reported several
times in the literature\cite{berkovits3,hik,hikami}.
We include the effects of absorption and frequency differences and find
\begin{mathletters} \label{hik}
\begin{eqnarray} H_4&=&h_4 [-{\bf q}_1
\cdot {\bf
q}_3 - {\bf q}_2 \cdot{\bf q}_4 + \frac{1}{2}\sum_{i=1}^4 ({\bf q}_i^2
+\kappa_i^2 +i\Omega_i) ] , \\ H_6& =&-h_6[{\bf
q}_1\cdot \!{\bf q}_2+{\bf q}_2 \cdot\!{\bf q}_3+
{\bf q}_3 \cdot\!{\bf q}_4+{\bf q}_4 \cdot\! {\bf q}_5
+{\bf q}_5 \cdot\!{\bf q}_6 +{\bf q}_6 \cdot\!{\bf q}_1 + \sum_{i=1}^6 ({\bf
q}_i^2+\frac{1}{2}\kappa_i^2+\frac{i}{2}\Omega_i)].
\end{eqnarray}
\end{mathletters}
We call these the ``invariant'' forms of $H_4$ and $H_6$ as they are
unchanged under the shifts ${\bf q}_i\rightarrow {\bf q}_i
-\frac{1}{4} \sum_{j=1}^4
{\bf q}_j$, ${\bf q}_i\rightarrow {\bf q}_i -\frac{1}{6}\sum_{j=1}^6 {\bf
q}_j$, respectively. We defined the pre-factors as \begin{equation} h_4
=\frac{\ell^5}{48
\pi k^2}, \qquad h_6=\frac{\ell^7}{96 \pi k^4}. \end{equation} The momenta of
the diffusons that are attached to these vertices are denoted by ${\bf q}_i$,
where the diffusons are numbered clockwise on the vertex and their momenta
are directed towards the vertex. In the actual calculations the Fourier
transforms in the $z$-direction of the vertices are used. Compared to
previous results of, for instance Hikami\cite{hikami}, the vertices contain
additional frequency and absorption terms. According to the diffusion
equation (\ref{difeq}) these extra terms together with the ${\bf q}^2$ terms,
lead to a source $\delta (z-z')$. For external diffusons, such terms are
neglected as they bring contributions of the order $\ell/L$. This
approximation simplified the calculation of for instance the long range
correlation function\cite{stephenboek}. For the internal diffusons, however,
the source terms are of leading order and cause divergencies. They correspond
to the situation where the two interferences take place within a distance of
a few mean free paths.

As an example we calculate the diagram presented in Fig.~\ref{figreg}(a)i.
This diagram was first depicted by Feng, Kane, Lee and Stone\cite{feng}
and considered in detail  by Berkovits and Feng\cite{berkovits3}. These
authors pointed out that a short distance divergency appears.
For the case of external momenta approximately zero,
the Hikami-box (\ref{hik}) yields  $H_4({\bf q},0,-{\bf q},0)=2h_4 q^2$,
while the internal diffuson
has the form ${\cal L}_{int}(q)=12\pi/(\ell^3 q^2)$. Omitting the external
lines the diagram Fig.~\ref{figreg}(a)i. then simply leads to
\begin{equation} \int \frac{ {\rm d}^3 q}{(2\pi)^3} H_4^2({\bf q},0,-{\bf
q},0) {\cal
L}_{int}^2(q) =\frac{\ell^4}{4k^4} \int \frac{ {\rm d}^3q}{(2\pi)^3} q^0=
\frac{\ell^4}{4k^4}\delta^{(3)}({\bf r}=0), \end{equation} which is  indeed a
cubic divergency in three dimensions, and, more generally, a $d$-dimensional
divergency in $d$ dimensions. As it is arising from the physically innocent
situation where the two interference vertices are close to each other, we
expect that the divergency has to disappear finally.

We now calculate the diagram for the slab geometry.
 For simplicity we first
consider a quasi one-dimensional system in which frequency
differences and absorption are
absent, therefore the decay rate (``mass'') vanishes,
i.e. $M=0$ for all diffusons. (Beyond quasi 1d one would have
to take non-zero $M=Q$ and sum over the allowed $Q$.)
{}From Fig.~\ref{figreg}(a)i one directly reads off its corresponding
expression $F_{a.i}$ \begin{equation}
F_{a.i} = \int \int {\rm d}z {\rm d} z'
{\cal L}_{\rm in}^2(z) H_4(z)
 {\cal L}_{int}^2(z,z') H_4 (z') {\cal L}_{\rm out}^2(z). \end{equation}
We label the two incoming diffusons 1 and 3, the outgoing ones 2 and 4, and
the internal ones ${\cal L}(z_5,z_7)$ and ${\cal L}(z_6,z_8)$, (with  $5,7$
at $z$ and $6,8$ at $z'$). The real space expressions for the
Hikami boxes become
\begin{mathletters}
\begin{eqnarray}
H_4(z)&=&h_4 [ \partial_{z_1} \partial_{z_3}
+\frac{1}{2}( \partial_{z_5} +\partial_{z_7})^2 - \partial_{z_5}^2-
\partial_{z_7}^2 ] , \\
H_4(z')&=&h_4 [ \partial_{z_2} \partial_{z_4}
+\frac{1}{2}( \partial_{z_6} +\partial_{z_8} )^2- \partial_{z_6}^2-
\partial_{z_8}^2 ]
,\end{eqnarray}
\end{mathletters}
 in which $\partial_{z_i}$ is the derivative of the
corresponding diffuson;
after performing the differentiation $z_{1,3,5,7}$ should be put equal to
$z$,
while $z_{2,4,6,8}$ should be put equal $z'$. Keeping $z_{2,4,6,8}$ fixed,
we  obtain for the $z$-integral after some partial integrations
\begin{eqnarray}
&&\int_0^L {\rm d} z H_4(z) {\cal L}_{\rm in}(z_1) {\cal L}_{\rm in}(z_3)
{\cal L}_{int}(z_5,z_6){\cal L}_{int}(z_7,z_8) \nonumber \\
&=&  \frac{12\pi h_4}{\ell^3}   \int {\rm d} z
 [{\cal L}_{int}(z,z_6) \delta(z\!-\!z_8)+
{\cal L}_{int}(z,z_8) \delta(z\!-\!z_6)]{\cal L}_{\rm in}^2(z) +2 h_4 \!
\int {\rm d} z
 {\cal L}_{int}(z,z_6){\cal L}_{int}(z,z_8) {\cal L}_{\rm
in}'^2(z)\nonumber \\  &=& \frac{12\pi h_4}{\ell^3}
{\cal L}_{int}(z_8,z_6){\cal L}_{\rm in}^2(z_8) +\frac{12\pi h_4}{\ell^3}
 {\cal L}_{int}(z_6,z_8){\cal L}_{\rm in}^2(z_6)+ 2h_4\int {\rm d} z
 {\cal L}_{int}(z,z_6){\cal L}_{int}(z,z_8) {\cal L}_{\rm in}'^2(z)
\nonumber . \end{eqnarray}
Here we also used the diffuson equation, which in this simplified case reads
$\partial_z^2 {\cal L}_{\rm in}=0$ and $\partial_z^2 {\cal
L}_{int}(z,z')=12\pi
\delta(z-z') /\ell^3$. Also carrying out the $z'$-integral we find
after performing again some partial integrations
\begin{eqnarray} F_{a.i}& =& \int  {\rm d} z' H_4(z') {\cal L}_{\rm out}(z_2)
{\cal L}_{\rm out}(z_4) \left[ \frac{12\pi h_4}{\ell^3}
{\cal L}_{int}(z_8,z_6){\cal L}_{\rm in}^2(z_8)  + \right. \nonumber \\
&& \left. \frac{12\pi h_4 }{\ell^3}{\cal L}_{int}(z_6,z_8){\cal L}_{\rm
in}^2(z_6) +
2h_4 \int {\rm d} z {\cal L}_{int}(z,z_6){\cal L}_{int}(z,z_8){\cal L}_{\rm
in}'(z) \right ] \nonumber \\
&=&
\frac{\ell^4}{4 k^4}\delta(0)
\int {\rm d}z {\cal L}_{\rm in}^2(z) {\cal L}_{\rm
out}^2(z) \nonumber \\&& + \frac{h_4 \ell^2}{k^2}\int  {\rm
d} z {\cal L}_{int}(z,z)\left[{\cal L}_{\rm in}'^2(z) {\cal L}_{\rm out}^2(z)
+{\cal L}_{\rm in}^2(z) {\cal L}_{\rm out}'^2(z)+ {\cal L}_{\rm in}'(z)
{\cal L}_{\rm out}'(z)
{\cal L}_{\rm in}(z) {\cal L}_{\rm out}(z) \right]  \nonumber \\
&& + 4 h_4^2\int {\rm d} z' \int  {\rm d} z {\cal L}_{int}^2(z,z'){\cal
L}_{\rm in}'^2(z) {\cal L}_{\rm out}'^2(z'). \end{eqnarray}
The spatial derivative of ${\cal L}(z)$ is denoted ${\cal L}'$.
All diffusons are simple linear functions in this case, yielding
\begin{equation}  F_{a.i}= \frac{2}{15} \delta(0)L +\frac{14}{45}.
\end{equation}
Note that the prefactors of the diffusons and the Hikami boxes have canceled
precisely. This is closely related to the universal character of conductance
fluctuations in electronic systems, see (\ref{C3-G}).

The term $\delta(0)$ is a linear divergency, which is the cause of all
troubles. In the three-dimensional case one has to take $Q\neq 0$. The
$\delta(0)L$ term will occur also for transversal momentum $Q\neq 0$, so that
the $Q$-sum yields the cubic divergency \begin{equation} W^2
\int\frac{{\rm d}^2 Q}{(2\pi)^2} \delta(0)L=W^2 L \delta^{(3)} (0),
\end{equation} as expected from the above bulk
consideration.

We now give the results of all diagrams of Fig.~\ref{figreg}. We no longer
restrict to the $M=0$ case. The expressions are labeled according to the
diagrams in figure, $F_a$, $F_b$ and $F_c$.
In the diagrams of
Fig.~\ref{figreg}(a), the decay rates of Eq.(\ref{eqdefm}) for the internal
diffusons are each other complex conjugate, $M$ and $M^\ast$. In the
diagrams
of Fig.~\ref{figreg}(b) both internal diffusons have the same decay rate.
Using
the definition of the Hikami vertices and the diffusion equation, we obtain
\begin{mathletters} \label{eqfdif}
\begin{eqnarray} F_{a}(M)&=&\frac{\ell^4}{2k^4} \delta(0)
\int {\rm d}z\,{\cal L}_{\rm in}^2  {\cal L}_{\rm out}^2 \nonumber \\ &&+
\frac{h_4 \ell^2}{2 k^2}{\rm Re} \int {\rm d}z\, {\cal
L}_{int}(z,z;M) \left[3{\cal L}_{\rm
in}'^2  {\cal L}_{\rm out}^2+3{\cal L}_{\rm in}^2{\cal L}_{\rm out}'^2
 +10 {\cal L}_{\rm in}'{\cal L}_{\rm out}'{\cal L}_{\rm in}
{\cal L}_{\rm out} -4i \Omega {\cal L}_{\rm in}^2{\cal L}_{\rm out}^2
 \right]
\nonumber \\&&
+4h_4^2\int \int \!{\rm d}z \,{\rm
d}z' \, {\cal L}_{int}(z,z';M) {\cal L}_{int}(z,z';M^\ast)\,
\left[ {\cal L}_{\rm in}'^2(z){\cal L}_{\rm out}'^2(z')+
{\cal L}_{\rm in}'(z) {\cal L}_{\rm out} '(z) {\cal L}_{\rm in}'(z')
{\cal L}_{\rm out} '(z') \right], \\
F_{b}(M)&=&
 \frac{\ell^4}{4k^4} \delta(0)
\int {\rm d}z{\cal L}_{\rm in}^2
 {\cal L}_{\rm out}^2 \nonumber \\ &&
+\frac{h_4 \ell^2}{2 k^2}{\rm Re} \int \! {\rm d}z\, {\cal L}_{int}(z,z;M)
\left[-2 {\cal L}_{\rm in}' {\cal L}_{\rm out}'{\cal L}_{\rm in}
{\cal L}_{\rm out} + {\cal L}_{\rm in}'^2 {\cal L}_{\rm out}^2+
{\cal L}_{\rm in}^2 {\cal L}_{\rm out}'^2-4\kappa^2{\cal L}_{\rm in}^2
 {\cal L}_{\rm out}^2 \right]
 \nonumber \\ &&
+\frac{1}{2}h_4^2\int \int \! {\rm d}z \,{\rm d}z' \,
\left[{\cal L}_{int}^2(z,z';M)+{\cal L}_{int}^2(z,z';M^\ast)\right]
\frac{d^2}{dz^2} \left[{\cal L}_{\rm in}(z){\cal L}_{\rm out}(z) \right]
\frac{d^2}{dz'^2} \left[{\cal L}_{\rm in}(z') {\cal L}_{\rm out}(z')\right]
  , \\
F_{c}(M) &=&-\frac{\ell^4}{ k^4 }
\delta(0)\int {\rm d}z\,{\cal L}_{\rm in}^2
  {\cal L}_{\rm out}^2 \nonumber \\ &&
-\frac{2h_4 \ell^2}{ k^2}{\rm Re}\int {\rm d}z \, {\cal L}_{int}(z,z;M)
[2 {\cal L}_{\rm in}' {\cal L}_{\rm out}' {\cal L}_{\rm in}
{\cal L}_{\rm out}+{\cal L}_{\rm in}'^2 {\cal L}_{\rm out}^2 +{\cal
L}_{\rm in}^2 {\cal L}_{\rm out}'^2 -(\kappa^2+i\Omega) {\cal L}_{\rm
in}^2 {\cal L}_{\rm out}^2 ],
\end{eqnarray}
\end{mathletters}
where we used the short hand notation that in the single integrals all
incoming and outgoing diffusons have argument $z$.
To obtain the
variance of the fluctuations in quasi-one dimension,  $F$'s is
evaluated at transverse momentum $Q=0$. The internal momentum $Q$ enters the
equations via the decay rate of the internal diffusons, defined by
$M^2=Q^2+\kappa^2+i\Omega$. In two and three dimensions a sum or
integral over the
transversal momentum has to be performed.

In (\ref{eqfdif}a,b) we
can distinguish  three contributions to $F$.
In the first term, both Hikami boxes operate on the internal diffusons,
yielding in the diffusion approximation a delta function evaluated in zero.
The resulting term is independent of all the momenta of the external
diffusons. It can be seen from the diffusion equation that a diffuson decays
rapidly if its momentum becomes large. Terms of the diffusons with few
scatterers are dominant at large momentum; they cause the divergence.
Our present description of these processes is incomplete.
In
order to see the cancellation of this divergency, calculation of the long
range
diagrams is not sufficient, so that the short distance processes have to be
examined in detail. This will be done in the next section.
The second term of (\ref{eqfdif}a,b) is a
single integral, it comes about when the boxes act on one internal and on one
external diffuson. This corresponds to the case where one internal diffuson
is almost empty, while the other diffuson contains a arbitrary number
of scatterers. In two and three dimensions the momentum integral diverges,
since for large $Q$ it behaves as $\int {\rm d}z \int {\rm d}^{d-1} Q {\cal
L}_{int}(Q;z,z) \sim \int {\rm d}^{d-1}Q \, Q^{-1}$. But when summing the
$a$, $b$ and $c$ contribution this term cancels.  The sum gives
\begin{eqnarray}\label{Fabc}
F_a(M)&&+F_b(M)+F_c(M)
=- \frac{\ell^4}{4 k^4}\delta(0)\int {\rm d}z{\cal L}_{\rm in}^2
 {\cal L}_{\rm out}^2 \nonumber \\ &&+
4 h_4^2\int \int \!{\rm d}z \,{\rm d}z' \,
{\cal L}_{int}(z,z';M) {\cal L}_{int}(z,z';M^\ast)\,
\left[ {\cal L}_{\rm in}'^2(z){\cal L}_{\rm out}'^2(z')+
{\cal L}_{\rm in}'(z) {\cal L}_{\rm out} '(z) {\cal L}_{\rm in}'(z')
{\cal L}_{\rm out} '(z') \right]\nonumber \\
&&+\frac{1}{2}h_4^2\int \int \!{\rm d}z \,{\rm d}z' \,
\left[{\cal L}_{int}^2(z,z';M)+{\cal
L}_{int}^2(z,z';M^\ast)\right]\frac{d^2}{dz^2}
\left[{\cal L}_{\rm in}(z){\cal L}_{\rm out}(z)\right]
\frac{d^2}{dz'^2}\left[{\cal L}_{\rm in}(z') {\cal L}_{\rm out}(z')
\right] \label{eqsumf}. \end{eqnarray}
The last two terms involve  a double integral describing interference
vertices
at different points in space; technically it arises when both boxes act
on external diffusons, or from terms where they do so after
partial integrations. This term is absent in the
expression $F_{c}$, which contains only one $z-$dependence as can be
seen from Fig.~\ref{figreg}(c). When performing the integral of $F$ over
the transverse momentum $Q$, the double integral term
behaves at large $Q$ as
$\int dz \int dz' \int {\rm d}^{d-1} Q {\cal L}_{int}^2
(Q;z,z')\sim \int {\rm d}^{d-1}Q \, Q^{-3}$. It is thus convergent.
One expects that finally only this term will survive. It is also the
only contribution depending solely on derivatives of the external diffusons.

In the quasi one-dimensional case where absorption and frequency terms are
absent, the expressions reduce to
\begin{equation} F_{a}(0)+F_b(0)+F_{c}(0) = -\frac{2}{15}\delta(0)L+
\frac{2}{15}. \end{equation}
The second part is a well known result for the UCF in one dimension, but a
singular part is annoyingly present. Before we can obtain the UCF and
correlation functions, we have to show its cancellation. We expect the
unphysical divergency to disappear by summing all diagrams and in the next
section we set out finding {\em all} leading diagrams.

\section{Generating diagrams through a generalized Kadanoff-Baym
technique}\label{secKB}
Finding the correct and complete set of diagrams by educated guess proves
very difficult, especially for the low order diagrams. In this section we
pursue the technique founded by Kadanoff
and Baym\cite{baym1,baym2} to generate all diagrams. This method provides a
systematic way to construct the diagrams in a particular approximation. The
approximation is made on the level of a generating functional. The form of
this functional is guided by intuition or by prior knowledge of the
self-energy or some other physical quantity. The theory of Kadanoff and Baym,
of which the basics will be reviewed in the course of this section,
prescribes which diagrams are to be included in the perturbation theory on
{\em any} level in the hierarchy of the many particle Green functions. It was
proven by Kadanof and Baym \cite{baym1,baym2} that this procedure provides a
conserving theory in the sense that sum-rules based on conservation of
particle number, momentum, angular momentum or energy are fulfilled. However,
it does not guarantee that a sensible theory in a physical or even in
mathematical sense will be produced. (The Green functions themselves may not
posses the correct analytical structure, i.e. they may be
non-Herglotz\cite{ruud_uhrig}).

The theory of Kadanoff and Baym is defined for interacting electron systems.
It is our goal to describe the scattering of light in a disordered medium.
The connection between electronic and optical disordered media can be put on
an solid footing as the equations which describe these phenomena can be
mapped onto each other\cite{vanhaeringen}. The equivalence of interacting
and disordered systems, however, is less obvious. Although one can rewrite
models of disorder in such a way that an effective interaction is
present\cite{efetov1},
 the precise form of that interaction
depends on the physical quantities that are studied. Furthermore, if one
wants to represent the disorder as a two-body, central potential, one is
limited to special disorder distributions (e.g. Gaussian) or approximations
(e.g. second order Born). However, these drawbacks do not hinder us at using
this theory in a heuristic sense.

\begin{figure}
\centerline{\psfig{file=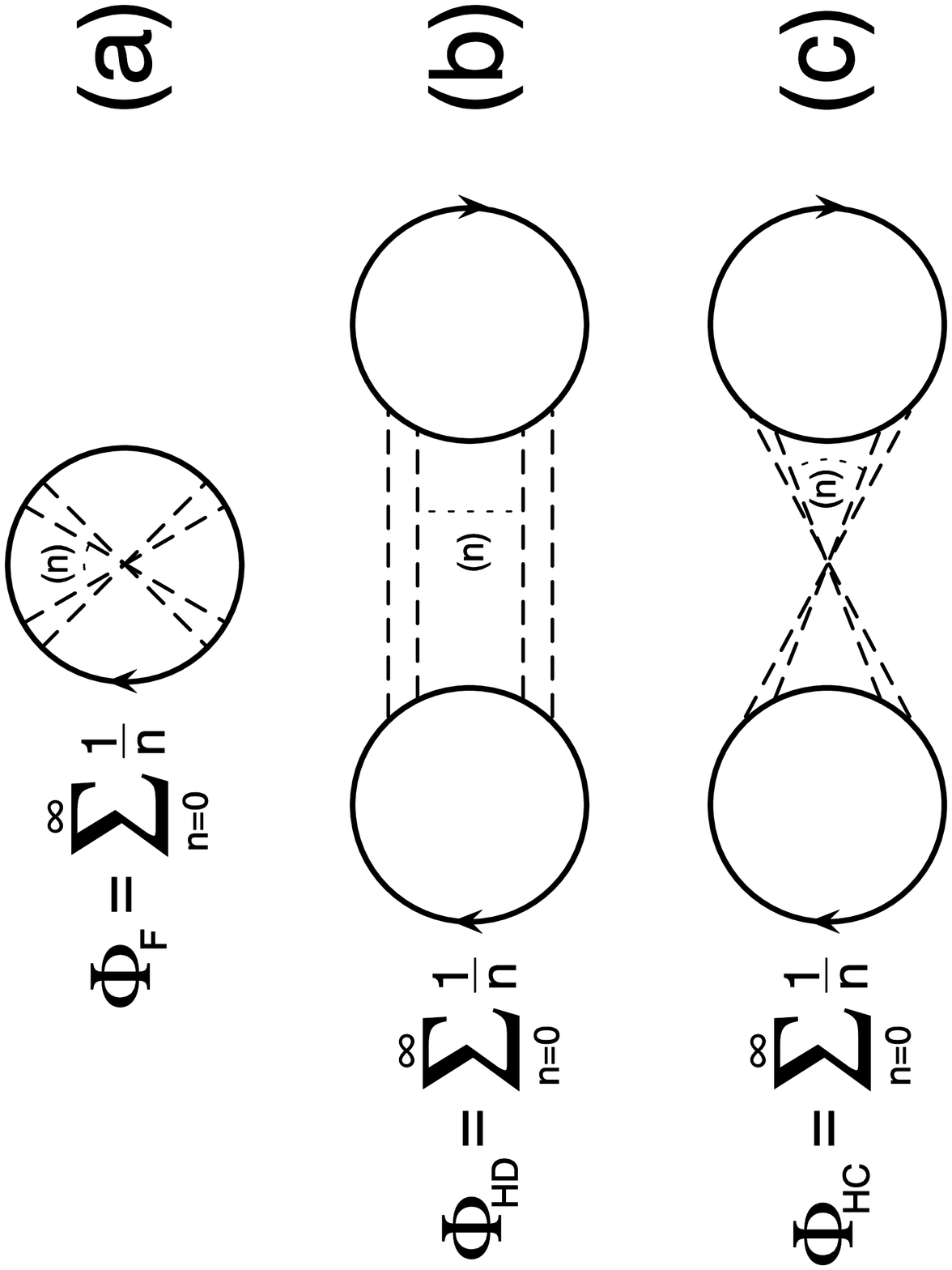,width=8cm}}
\caption{The generating functional which produces diagrams for (a)
      weak localization, (b) and (c) the UCF.}
     \label{fi:r1}
\end{figure}

The input of the theory is an approximate functional $\Phi$ from which we may
derive the perturbation theory. In the choice we are lead by the theory of
weak localization and universal conductance fluctuations (UCF). The maximally
crossed diagrams, which are the most important quantum corrections, are
generated by the functional depicted in Fig.\ \ref{fi:r1}(a). These
wheel-like diagrams are generalized Fock diagrams in the sense that the first
term in the sum produces the Fock diagram for the self-energy. The UCF
diagrams are generated by a functional which may be called "generalized
Hartree" in the same sense, see Fig.\ \ref{fi:r1}(b) and \ref{fi:r1}(c). It
turns out that the weak-localization effect and the UCF are connected and may
be generated from a single diagram\cite{ruud_tobe}. This observation leads
to the conjecture that the most important contributions in the intensity
fluctuations stem from the same diagrams. We thus adopt the functionals as
given in Fig.\ \ref{fi:r1}(b) and \ref{fi:r1}(c), as the ones we start our
calculation with.

Let us shortly review the aspects of the Kadanoff and Baym theory we need.
Starting with some functional $\Phi$ as described above, the self-energy
$\Sigma_{\scriptscriptstyle \bar1}^{\scriptscriptstyle1}\,$ and the
two-particle irreducible vertex
$\Sigma_{\scriptscriptstyle\bar1\bar2}^{\scriptscriptstyle 12}$
may be found from \begin{mathletters} \label{defsig} \begin{eqnarray}
\Sigma_{\scriptscriptstyle\bar1}^{\scriptscriptstyle1} \,\!&=&{\delta\Phi}\;
/ \; \delta G_{\scriptscriptstyle1}^{\scriptscriptstyle\bar 1}\, \\
\Sigma_{\scriptscriptstyle\bar1\bar2}^{\scriptscriptstyle12}
\,\!&=&{\delta^2\Phi}\; / \;(\delta G_{\scriptscriptstyle
1}^{\scriptscriptstyle \bar 1}\,\,\delta G_{\scriptscriptstyle
2}^{\scriptscriptstyle \bar 2}), \end{eqnarray}  \end{mathletters}
where the numbers $1,\bar1,2, \cdots$ denote the collection of variables
$(\vec{k}_1,\omega_1)$, $(\vec{k}_1',\omega_1')$,
$(\vec{k}_2,\omega_2)\cdots$
and $G_{\scriptscriptstyle 1}^{\scriptscriptstyle \bar 1}=-\langle{\cal
T}\Psi(1)\Psi^+(\bar1)\rangle$. At this point we generalize these results to
obtain irreducible three- and four-particle vertices
\begin{mathletters} \label{deftet}  \begin{eqnarray}
 \Sigma_{\scriptscriptstyle\bar1\bar2 \bar 3}^{
\scriptscriptstyle123}\, \!&=&{\delta^3\Phi}\;
/ \;  (\delta G_{\scriptscriptstyle 1}^{\scriptscriptstyle \bar  1}\,\,
\delta G_{\scriptscriptstyle 2}^{\scriptscriptstyle \bar  2}\,\, \delta
G_{\scriptscriptstyle 3}^{\scriptscriptstyle \bar  3}) \\
 \Sigma_{\scriptscriptstyle\bar1\bar2 \bar3\bar4}^{
\scriptscriptstyle1234}\,\!&=&{\delta^4\Phi}\;
/ \;  (\delta G_{\scriptscriptstyle 1}^{\scriptscriptstyle \bar  1}\,\,
\delta G_{\scriptscriptstyle 2}^{\scriptscriptstyle \bar  2}\,\, \delta
G_{\scriptscriptstyle 3}^{\scriptscriptstyle \bar  3}\,\, \delta
G_{\scriptscriptstyle 4}^{\scriptscriptstyle \bar  4}).
\end{eqnarray}  \end{mathletters}
Note that these generalizations, as the ones that will follow,  are
not free from internal symmetry. Since we are primarily interested
in the structure of the diagrams, rather
then their multiplicativity, we will make no effort to remove this symmetry.
The one-particle Green function
$G_{\scriptscriptstyle1}^{\scriptscriptstyle\bar1}$
is connected to the self-energy by the Dyson
equation. A two-particle Green function
$G_{\scriptscriptstyle12}^{\scriptscriptstyle\bar1\bar2}$ is connected to
the irreducible two-particle vertex by the Bethe-Salpheter equation
\begin{mathletters} \label{defgreen}\begin{eqnarray}
  G_{\scriptscriptstyle 1}^{\scriptscriptstyle \bar
1}&=&G_{\scriptscriptstyle 1}^{\scriptscriptstyle \bar
1}(0)+G_{\scriptscriptstyle 1}^{\scriptscriptstyle \bar
2}(0)\,\Sigma_{\scriptscriptstyle
\bar2}^{\scriptscriptstyle2}\,G_{\scriptscriptstyle 2}^{\scriptscriptstyle
\bar 1} \\
G_{\scriptscriptstyle12}^{\scriptscriptstyle\bar
1\bar2}&=&G_{\scriptscriptstyle 1}^{\scriptscriptstyle \bar
2}\,G_{\scriptscriptstyle 2}^{\scriptscriptstyle \bar
1}+G_{\scriptscriptstyle 1}^{\scriptscriptstyle \bar  3
}\,G_{\scriptscriptstyle 3}^{\scriptscriptstyle \bar
1}\,\Sigma_{\scriptscriptstyle\bar3\bar4}^{
\scriptscriptstyle34 }\, G_{\scriptscriptstyle42}^{\scriptscriptstyle\bar
4\bar2}\, , \end{eqnarray} \end{mathletters}
where $G_{\scriptscriptstyle 1}^{\scriptscriptstyle \bar  1}(0)$ indicates
the bare propagator. Integration over
the free variables occurring as super and subscript is assumed.
Again, we may generalize these  results to obtain a three-particle
Green function
$G_{\scriptscriptstyle123}^{\scriptscriptstyle \bar1\bar2\bar3}$
and a four-particle Green function
$G_{\scriptscriptstyle1234}^{\scriptscriptstyle\bar1\bar2\bar3\bar4}$. We
find \cite{ruud_tobe} \begin{mathletters} \label{defdrie}  \begin{eqnarray}
  G_{\scriptscriptstyle123}^{\scriptscriptstyle\bar1\bar2\bar3}&=&
(G_{\scriptscriptstyle 5}^{\scriptscriptstyle \bar
1}\,G_{\scriptscriptstyle13}^{\scriptscriptstyle\bar5\bar3}\,
 + G_{\scriptscriptstyle 1}^{\scriptscriptstyle \bar
5}\,G_{\scriptscriptstyle53}^{\scriptscriptstyle\bar1\bar3}\,\!) (\delta_{
\scriptscriptstyle2}^{ \scriptscriptstyle5}\,
\delta_{ \scriptscriptstyle \bar5}^{ \scriptscriptstyle\bar 2}  +
\Sigma_{\scriptscriptstyle\bar4\bar5}^{
\scriptscriptstyle45}\,
G_{\scriptscriptstyle24}^{\scriptscriptstyle\bar2\bar4}\,)  +
  G_{\scriptscriptstyle 4}^{\scriptscriptstyle \bar 1} G_{\scriptscriptstyle
1}^{\scriptscriptstyle \bar 4}\, (\Sigma_{\scriptscriptstyle\bar4 \bar5}^{
\scriptscriptstyle45}\,
G_{\scriptscriptstyle235}^{\scriptscriptstyle\bar2\bar3\bar5}\,
\!+ \Sigma_{\scriptscriptstyle\bar4\bar 5\bar 6}^{ \scriptscriptstyle456}\,
G_{\scriptscriptstyle36 }^{ \scriptscriptstyle\bar3\bar6}\,
G_{\scriptscriptstyle25}^{\scriptscriptstyle\bar 2\bar5}\, \!) \\
  G_{\scriptscriptstyle1234}^{ \scriptscriptstyle \bar1\bar2\bar3\bar4} &=&
(G_{\scriptscriptstyle534}^{\scriptscriptstyle\bar1\bar3\bar4}\,
G_{\scriptscriptstyle 1}^{\scriptscriptstyle \bar 5} \, \!+
G_{\scriptscriptstyle 53}^{\scriptscriptstyle \bar 1\bar3}\,
G_{\scriptscriptstyle 14}^{\scriptscriptstyle\bar5\bar4}\, \!+
G_{\scriptscriptstyle 54}^{\scriptscriptstyle\bar1\bar4}\,
G_{\scriptscriptstyle 13}^{\scriptscriptstyle\bar5\bar3}\,  \!+
G_{\scriptscriptstyle 5}^{\scriptscriptstyle \bar3}\,
G_{\scriptscriptstyle 134}^{\scriptscriptstyle\bar5\bar3\bar 4}\,\!)
(\delta_{ \scriptscriptstyle2}^{ \scriptscriptstyle5}\,\delta_{
\scriptscriptstyle \bar5}^{
\scriptscriptstyle\bar2}\, \!+ \Sigma_{\scriptscriptstyle\bar5\bar 6}^{
\scriptscriptstyle56}\, G_{\scriptscriptstyle26}^{
\scriptscriptstyle\bar2\bar6}\, \!) + \nonumber \\   &
&(G_{\scriptscriptstyle58}^{\scriptscriptstyle\bar 1\bar8}\,
G_{\scriptscriptstyle 1}^{\scriptscriptstyle \bar 5}\, \!+
G_{\scriptscriptstyle 5}^{\scriptscriptstyle \bar 1}\,
G_{\scriptscriptstyle18}^{\scriptscriptstyle\bar5\bar8}\,  \!) (
\Sigma_{\scriptscriptstyle\bar5\bar6\bar 7}^{ \scriptscriptstyle567}\,
G_{\scriptscriptstyle 79}^{\scriptscriptstyle\bar 7\bar9}\,
G_{\scriptscriptstyle26}^{\scriptscriptstyle \bar2\bar6}\,
\!+ \Sigma_{\scriptscriptstyle\bar5\bar6 }^{
\scriptscriptstyle56}\,G_{\scriptscriptstyle269}^{ \scriptscriptstyle
\bar2\bar6\bar9
}\,\!) (\delta_{ \scriptscriptstyle3}^{ \scriptscriptstyle8}\,\delta_{
\scriptscriptstyle
\bar8}^{ \scriptscriptstyle\bar3}\,\delta_{ \scriptscriptstyle4}^{
\scriptscriptstyle9}\,\delta_{ \scriptscriptstyle \bar9}^{
\scriptscriptstyle\bar4}  + \delta_{ \scriptscriptstyle4}^{
\scriptscriptstyle8}\,\delta_{ \scriptscriptstyle
\bar8}^{ \scriptscriptstyle\bar4}\,\delta_{ \scriptscriptstyle\bar3}^{
\scriptscriptstyle9}\,\delta_{ \scriptscriptstyle \bar9}^{
\scriptscriptstyle\bar3} ) + \nonumber \\   & &G_{\scriptscriptstyle
5}^{\scriptscriptstyle
\bar 1}\,G_{\scriptscriptstyle 1}^{\scriptscriptstyle \bar
5}\left(\Sigma_{\scriptscriptstyle\bar5\bar6
\bar7\bar8}^{ \scriptscriptstyle5678}\,G_{\scriptscriptstyle48
}^{\scriptscriptstyle\bar4\bar8}\,G_{\scriptscriptstyle37
}^{\scriptscriptstyle \bar3\bar7}\,
G_{\scriptscriptstyle26}^{ \scriptscriptstyle\bar2\bar6}
 +
\Sigma_{\scriptscriptstyle\bar5\bar6\bar7}^{
\scriptscriptstyle567}\,(G_{\scriptscriptstyle347}^{ \scriptscriptstyle
\bar3\bar4\bar7}\,G_{ \scriptscriptstyle26}^{\scriptscriptstyle
\bar2\bar6}\,
\!+ G_{\scriptscriptstyle37}^{\scriptscriptstyle\bar3\bar7}\,
G_{\scriptscriptstyle246}^{ \scriptscriptstyle \bar2\bar4\bar6}\,
  \!+ G_{\scriptscriptstyle47}^{\scriptscriptstyle\bar4\bar
7}G_{\scriptscriptstyle236}^{ \scriptscriptstyle \bar2\bar3\bar6}\,\!) +
\Sigma_{\scriptscriptstyle\bar5\bar6 }^{
\scriptscriptstyle56}\, G_{\scriptscriptstyle2346}^{ \scriptscriptstyle
\bar2\bar3\bar4\bar6}\,\!\right) \label{defdrieb} , \end{eqnarray}
\end{mathletters} where the $\delta$'s are Kronecker delta's.
These equations provide the full information on the fluctuations in
transport quantities given the exact set of irreducible vertices
$\Sigma_{\scriptscriptstyle
\bar1}^{\scriptscriptstyle1}\,
,\Sigma_{\scriptscriptstyle\bar1\bar 2}^{
\scriptscriptstyle12}\,
,\Sigma_{\scriptscriptstyle\bar1\bar2 \bar 3}^{
\scriptscriptstyle123}\,,\Sigma_{\scriptscriptstyle\bar1\bar2\bar3
\bar4}^{ \scriptscriptstyle1234}$. An approximation of this
set can be found by choosing a suitable functional $\Phi$.

The functional we adopt is
\begin{mathletters} \label{deffunc}\begin{eqnarray}
   \Phi&:=& G_{\scriptscriptstyle 1}^{\scriptscriptstyle \bar
1}\,G_{\scriptscriptstyle 2}^{\scriptscriptstyle \bar  2}\left(-
S_{\scriptscriptstyle \bar1\bar2}^{ \scriptscriptstyle12} +
\sum_{n=1}^{\infty}    {\scriptstyle\frac{1}{n}}L_{\scriptscriptstyle \bar
1\bar2}^{\scriptscriptstyle12} [n] +  {\scriptstyle\frac{1}{n}}
C_{\scriptscriptstyle \bar 1\bar2}^{\scriptscriptstyle12} [n]\right),
\end{eqnarray} \end{mathletters}
where $L_{\scriptscriptstyle \bar 1\bar2}^{\scriptscriptstyle12}[n]$ and
$C_{\scriptscriptstyle \bar 1\bar 2}^{\scriptscriptstyle 12}[n]$ are the
product of $n$ scattering terms in the particle-hole
and particle-particle channel respectively. The last two terms are
the ones depicted
in Fig.\ref{fi:r1}(b) and \ref{fi:r1}(c). Algebraicly we have
\newcommand{\rv}{{{\makebox[0pt][l]
{$\scriptscriptstyle\hspace{0.4pt}\cdot$}}\phantom{0}}}
\begin{mathletters} \label{deflads}\begin{eqnarray}
  S_{\scriptscriptstyle \bar1\bar2}^{ \scriptscriptstyle12}  &=&
L_{\scriptscriptstyle \bar 1\bar2}^{\scriptscriptstyle12}\,[1] =
C_{\scriptscriptstyle \bar1 \bar2}^{\scriptscriptstyle12}\,[1] =
\delta_{\scriptscriptstyle\bar2}^{\scriptscriptstyle1}\,\Gamma\,
\delta_{\scriptscriptstyle\bar1}^{\scriptscriptstyle2}
\\   L_{\scriptscriptstyle \bar 1\bar2}^{\scriptscriptstyle12} [n] &=&
S_{\scriptscriptstyle
\bar1\bar3}^{ \scriptscriptstyle13}\,G_{\scriptscriptstyle
4}^{\scriptscriptstyle
\bar 3}\,G_{\scriptscriptstyle 3}^{\scriptscriptstyle \bar
4}\,S_{\scriptscriptstyle \bar4\bar{\rv}}^{ \scriptscriptstyle4{\rv}} \cdots
S_{\scriptscriptstyle\bar{\rv}\bar2}^{ \scriptscriptstyle{\rv}2} \\
  C_{\scriptscriptstyle \bar1 \bar2}^{\scriptscriptstyle12} [n] &=&
S_{\scriptscriptstyle
\bar1\bar2}^{ \scriptscriptstyle34}\,G_{\scriptscriptstyle
4}^{\scriptscriptstyle
\bar 3}\,G_{\scriptscriptstyle 3}^{\scriptscriptstyle \bar
4}\,S_{\scriptscriptstyle\bar 4\bar 3}^{ \scriptscriptstyle
{\rv}{\rv}}\cdots
 S_{\scriptscriptstyle\bar{\rv}\bar{\rv}}^{ \scriptscriptstyle44},
\end{eqnarray} \end{mathletters}
where $\Gamma$ is the scattering strength in second order Born approximation.
After differentiation of $\Phi$ with respect to $G$ ladder diagrams and
maximally crossed diagrams
appear. Each element in these series have prefactor one. Therefore we
 define the
full ladders in both channels as
\begin{mathletters} \label{diffusons}  \begin{eqnarray}
   L_{\scriptscriptstyle \bar 1\bar2}^{\scriptscriptstyle12} &=&
S_{\scriptscriptstyle
\bar1\bar2}^{ \scriptscriptstyle12}  + S_{\scriptscriptstyle \bar1\bar3}^{
\scriptscriptstyle13}\,G_{\scriptscriptstyle
4}^{\scriptscriptstyle \bar  3}\,G_{\scriptscriptstyle
3}^{\scriptscriptstyle \bar
4}\,L_{\scriptscriptstyle \bar 4\bar 2}^{\scriptscriptstyle42} =
\sum_{n=1}^{\infty}
L_{\scriptscriptstyle \bar 1\bar 2}^{\scriptscriptstyle12} [n]{}  \\
   C_{\scriptscriptstyle \bar1 \bar2}^{\scriptscriptstyle12} &=&
S_{\scriptscriptstyle \bar1\bar2}^{ \scriptscriptstyle12}
 + S_{\scriptscriptstyle\bar1\bar2}^{ \scriptscriptstyle 43}
\,G_{\scriptscriptstyle
4}^{\scriptscriptstyle \bar  3}\,G_{\scriptscriptstyle
3}^{\scriptscriptstyle \bar  4
}\,C_{\scriptscriptstyle \bar4 \bar3}^{\scriptscriptstyle12} =
\sum_{n=1}^{\infty}
C_{\scriptscriptstyle  \bar1 \bar2}^{\scriptscriptstyle12} [n]{}.
\end{eqnarray} \end{mathletters}

Systematic application of Eqs.(\ref{defsig}) and (\ref{deftet}) on the
functional
(\ref{deffunc}) leads to the following irreducible $n$-particle vertices
\begin{mathletters} \label{npartic}\begin{eqnarray}
  \Sigma_{\scriptscriptstyle \bar1}^{\scriptscriptstyle1} \, &=&
G_{\scriptscriptstyle
2}^{\scriptscriptstyle \bar  2}\, (L_{\scriptscriptstyle \bar
2\bar1}^{\scriptscriptstyle12}  +
C_{\scriptscriptstyle \bar2 \bar1}^{\scriptscriptstyle12}  -
S_{\scriptscriptstyle\bar2\bar1}^{ \scriptscriptstyle12} ) \\
  \Sigma_{\scriptscriptstyle\bar1\bar2 }^{ \scriptscriptstyle12}\,  &=&
L_{\scriptscriptstyle  \bar 2\bar1}^{\scriptscriptstyle12}  +
C_{\scriptscriptstyle \bar2 \bar1}^{\scriptscriptstyle12}  -
S_{\scriptscriptstyle \bar2\bar1}^{ \scriptscriptstyle12} +
   G_{\scriptscriptstyle 3}^{\scriptscriptstyle \bar
3}\,G_{\scriptscriptstyle
4}^{\scriptscriptstyle \bar  4}\, (L_{\scriptscriptstyle
\bar1\bar4}^{\scriptscriptstyle32}\,L_{\scriptscriptstyle
\bar3\bar2}^{\scriptscriptstyle
14}  + L_{\scriptscriptstyle \bar1\bar2}^{\scriptscriptstyle
34}\,L_{\scriptscriptstyle
\bar3\bar4}^{\scriptscriptstyle 12}  + C_{\scriptscriptstyle \bar1
\bar4}^{\scriptscriptstyle32}\,C_{\scriptscriptstyle \bar3
\bar2}^{\scriptscriptstyle14} + C_{\scriptscriptstyle \bar1
\bar4}^{\scriptscriptstyle23}\, C_{\scriptscriptstyle \bar2
\bar3}^{\scriptscriptstyle14} ) \\
  \Sigma_{\scriptscriptstyle \bar1\bar2\bar3}^{ \scriptscriptstyle123}\, &=&
\mathop{\rm
Per}\nolimits_{\scriptscriptstyle\bar1\bar2\bar3}^{\scriptscriptstyle
123}\,\left\{
G_{\scriptscriptstyle 5}^{\scriptscriptstyle \bar
5}\,(L_{\scriptscriptstyle
\bar1\bar2}^{\scriptscriptstyle 35}\, L_{\scriptscriptstyle \bar
3\bar5}^{\scriptscriptstyle12}  +C_{\scriptscriptstyle \bar1
\bar2}^{\scriptscriptstyle35}\,C_{\scriptscriptstyle \bar3
\bar5}^{\scriptscriptstyle12} )\; + \right.
\nonumber \\    & & \left. G_{\scriptscriptstyle 5}^{\scriptscriptstyle \bar
5}\,G_{\scriptscriptstyle 6}^{\scriptscriptstyle \bar
6}\,G_{\scriptscriptstyle 7}^{\scriptscriptstyle \bar  7}\,
\left({\scriptscriptstyle\frac{1}{3}}L_{\scriptscriptstyle
\bar5\bar2}^{\scriptscriptstyle
16} \,L_{\scriptscriptstyle \bar
1\bar3}^{\scriptscriptstyle57}\, L_{\scriptscriptstyle
\bar3\bar 6}^{\scriptscriptstyle 72}    +
L_{\scriptscriptstyle \bar5\bar2}^{\scriptscriptstyle 16}
\, L_{\scriptscriptstyle
\bar1\bar3}^{\scriptscriptstyle 57}\, L_{\scriptscriptstyle
\bar7\bar6}^{\scriptscriptstyle32}  +
{\scriptscriptstyle\frac{1}{3}}C_{\scriptscriptstyle \bar5
\bar2}^{\scriptscriptstyle16}\,C_{\scriptscriptstyle \bar1
\bar6}^{\scriptscriptstyle73}\, C_{\scriptscriptstyle \bar3
\bar7}^{\scriptscriptstyle52}  + C_{\scriptscriptstyle \bar5
\bar2}^{\scriptscriptstyle16}\,C_{\scriptscriptstyle \bar1
\bar6}^{\scriptscriptstyle37}\, C_{\scriptscriptstyle \bar7
\bar3}^{\scriptscriptstyle52} \right)\right\}
\\   \Sigma_{\scriptscriptstyle\bar1\bar2\bar3\bar4}^{
\scriptscriptstyle1234}\, &=&
\mathop{\rm
Per}\nolimits_{\scriptscriptstyle\bar1\bar2\bar3\bar4}^{
\scriptscriptstyle1234}\,\left\{
{\scriptscriptstyle\frac{1}{4}}  (L_{\scriptscriptstyle
\bar1\bar2}^{\scriptscriptstyle
34}\, L_{\scriptscriptstyle \bar3\bar4}^{\scriptscriptstyle 12}  +
C_{\scriptscriptstyle
\bar1 \bar2}^{\scriptscriptstyle34} \, C_{\scriptscriptstyle \bar3
\bar4}^{\scriptscriptstyle12}) \, +
\right. \nonumber \\ & & G_{\scriptscriptstyle 5}^{\scriptscriptstyle \bar
5}\, G_{\scriptscriptstyle 6}^{\scriptscriptstyle \bar  6} \,
(L_{\scriptscriptstyle \bar1\bar4}^{\scriptscriptstyle
35}\, L_{\scriptscriptstyle \bar
6\bar2}^{\scriptscriptstyle45}\, L_{\scriptscriptstyle
\bar3\bar5}^{\scriptscriptstyle
12}  + {\scriptscriptstyle\frac{1}{2}} L_{\scriptscriptstyle
\bar1\bar6}^{\scriptscriptstyle
34}\, L_{\scriptscriptstyle \bar4\bar2}^{\scriptscriptstyle
65} \, L_{\scriptscriptstyle \bar3\bar5}^{\scriptscriptstyle 12}  +
{\scriptscriptstyle\frac{1}{2}}L_{\scriptscriptstyle \bar 1\bar2
}^{\scriptscriptstyle35}\,L_{\scriptscriptstyle \bar
3\bar4}^{\scriptscriptstyle16}\,L_{\scriptscriptstyle
\bar6\bar5}^{\scriptscriptstyle
42}  +   C_{\scriptscriptstyle \bar1
\bar2}^{\scriptscriptstyle46}\,C_{\scriptscriptstyle
\bar6 \bar4}^{\scriptscriptstyle 35}\,C_{\scriptscriptstyle \bar3
\bar5}^{\scriptscriptstyle12}  + C_{\scriptscriptstyle \bar1
\bar2}^{\scriptscriptstyle64}\, C_{\scriptscriptstyle \bar4
\bar6}^{\scriptscriptstyle35}\, C_{\scriptscriptstyle \bar3
\bar5}^{\scriptscriptstyle12} )
\,+ \nonumber \\   & & G_{\scriptscriptstyle 5}^{\scriptscriptstyle \bar
5}\, G_{\scriptscriptstyle 6}^{\scriptscriptstyle \bar
6}\, G_{\scriptscriptstyle 7}^{\scriptscriptstyle \bar
7}\, G_{\scriptscriptstyle 8}^{\scriptscriptstyle \bar  8}
( {\scriptscriptstyle \frac{1}{6}}L_{\scriptscriptstyle
\bar1\bar8}^{\scriptscriptstyle
54}\, L_{\scriptscriptstyle \bar4\bar7}^{\scriptscriptstyle
83}L_{\scriptscriptstyle
\bar3\bar6}^{\scriptscriptstyle 72}\, L_{\scriptscriptstyle
\bar5\bar2}^{\scriptscriptstyle 16} + L_{\scriptscriptstyle
\bar1\bar4}^{\scriptscriptstyle 58}\, L_{\scriptscriptstyle
\bar8\bar7}^{\scriptscriptstyle 43}\, L_{\scriptscriptstyle \bar
3\bar6}^{\scriptscriptstyle 72}\,  L_{\scriptscriptstyle
\bar5\bar2}^{\scriptscriptstyle
16}  +    {\scriptscriptstyle\frac{1}{2}}L_{\scriptscriptstyle
\bar1\bar4}^{\scriptscriptstyle 58}\, L_{\scriptscriptstyle
\bar8\bar3}^{\scriptscriptstyle 47}\, L_{\scriptscriptstyle
\bar7\bar6}^{\scriptscriptstyle 32}
\, L_{\scriptscriptstyle \bar5\bar2}^{\scriptscriptstyle 16}  +
{\scriptscriptstyle \frac{1}{2}}L_{\scriptscriptstyle
\bar1\bar3}^{\scriptscriptstyle
57}\, L_{\scriptscriptstyle \bar7\bar6}^{\scriptscriptstyle 32}
\, L_{\scriptscriptstyle
\bar5\bar8}^{\scriptscriptstyle 14}\, L_{\scriptscriptstyle
\bar4\bar2}^{\scriptscriptstyle 86}  + \nonumber
\\ & & \left.    {\scriptstyle\frac{1}{6} }C_{\scriptscriptstyle \bar1
\bar6}^{\scriptscriptstyle84}\,C_{\scriptscriptstyle \bar4
\bar8}^{\scriptscriptstyle73}\,C_{\scriptscriptstyle \bar3
\bar7}^{\scriptscriptstyle52}\,C_{\scriptscriptstyle \bar5
\bar2}^{\scriptscriptstyle16}  + C_{\scriptscriptstyle \bar1
\bar6}^{\scriptscriptstyle48}\,C_{\scriptscriptstyle \bar8
\bar4}^{\scriptscriptstyle73}\,C_{\scriptscriptstyle \bar3
\bar7}^{\scriptscriptstyle52}\,C_{\scriptscriptstyle \bar5
\bar2}^{\scriptscriptstyle16}
 + {\scriptstyle\frac{1}{2}}C_{\scriptscriptstyle \bar1
\bar6}^{\scriptscriptstyle48}\,C_{\scriptscriptstyle \bar8
\bar4}^{\scriptscriptstyle37}\,C_{\scriptscriptstyle \bar7
\bar3}^{\scriptscriptstyle52}\,C_{\scriptscriptstyle \bar5
\bar2}^{\scriptscriptstyle16}
 +   {\scriptstyle\frac{1}{4}}C_{\scriptscriptstyle \bar1
\bar6}^{\scriptscriptstyle37}\,C_{\scriptscriptstyle \bar7
\bar3}^{\scriptscriptstyle52}\,C_{\scriptscriptstyle \bar5
\bar2}^{\scriptscriptstyle84}\,C_{\scriptscriptstyle \bar4
\bar8}^{\scriptscriptstyle16})\right\}, \end{eqnarray} \end{mathletters}
where the operator $\mathop{\rm
Per}\nolimits_{\scriptscriptstyle\bar1\bar2\cdots \bar
n}^{\scriptscriptstyle 12\cdots
n}$ produces all the permutations of its operand in the variables
$(1,\bar1),(2,\bar2),\cdots,(n,\bar n)$, i.e. $n!$ diagrams. Expansion of
Eq.(\ref{defdrie})  using Eq.(\ref{npartic}) is  obviously a considerable
task. Taking into account that we only need those diagrams which are
different in a topological sense, the effort stays manageable. As usual,
diagrams which contain loops do not contribute.

\begin{figure}
\centerline{\psfig{file=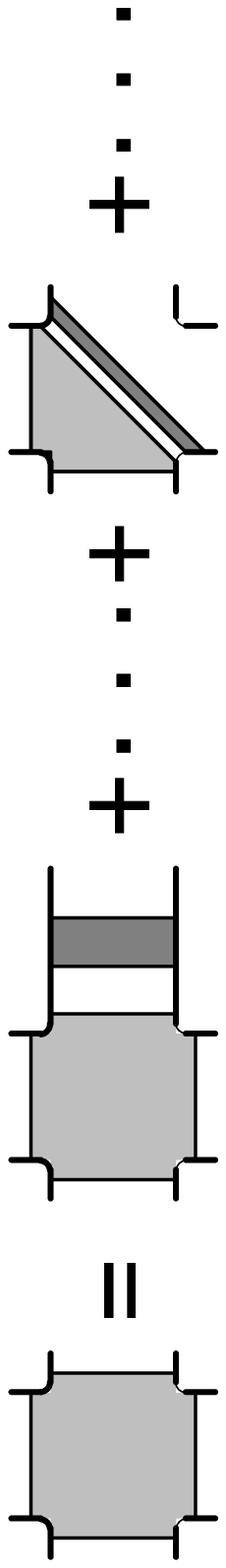,width=12cm}}
\caption{Part of
the integral equation from Eq.(\protect{\ref{defdrieb}}). First
indicated term on the       r.h.s. produces the external ladders.}
     \label{fi:r2}
\end{figure}

\begin{figure}
\centerline{\psfig{file=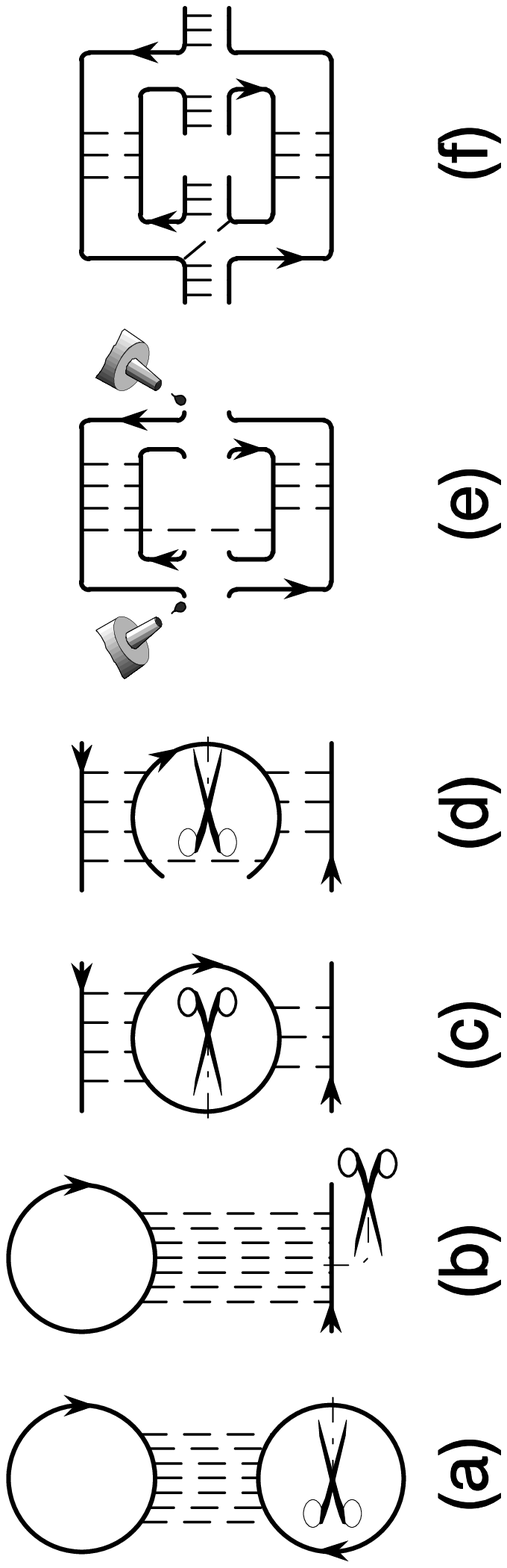,width=10cm}}
\caption{Particular example how the double diffuson structure is produced.
 Functional
     differention with respect to G is indicated by cutting out a line (pair
     of scissors) from (a) to (e). Pasting the result in the integral
equation,
     indicated in Fig.\ \protect{\ref{fi:r1}}, results in adding external
ladders (glue). }      \label{fi:r3}
\end{figure}

To illustrate the procedure set out above we have a closer look at two
 examples.
First of all, part of the integral equation for
$G_{\scriptscriptstyle1234}^{ \scriptscriptstyle \bar1\bar2\bar3\bar4}$
is needed. The relevant part is depicted in Fig.\ \ref{fi:r2}. Then, in
Fig.\ \ref{fi:r3}(a) we start off with the functional as depicted in Fig.\
\ref{fi:r1}(b). Successive functional differentiation (cutting) as
prescribed in
Eqs.(\ref{defsig}) and (\ref{deftet}) produce the one-, two-, three- and
four-particle irreducible vertices, given in Figs\ \ref{fi:r3}(b),
\ref{fi:r3}(c), \ref{fi:r3}(d) and \ref{fi:r3}(e) respectively.  Insertion
(glueing) of
the latter irreducible four-particle vertex in Fig.\ \ref{fi:r2} amounts to
adding external diffusons. Of course, more diagrams are produced in this
particular example but for the sake of simplicity no attention is paid to
these diagrams.

\begin{figure}
 \centerline{\psfig{file=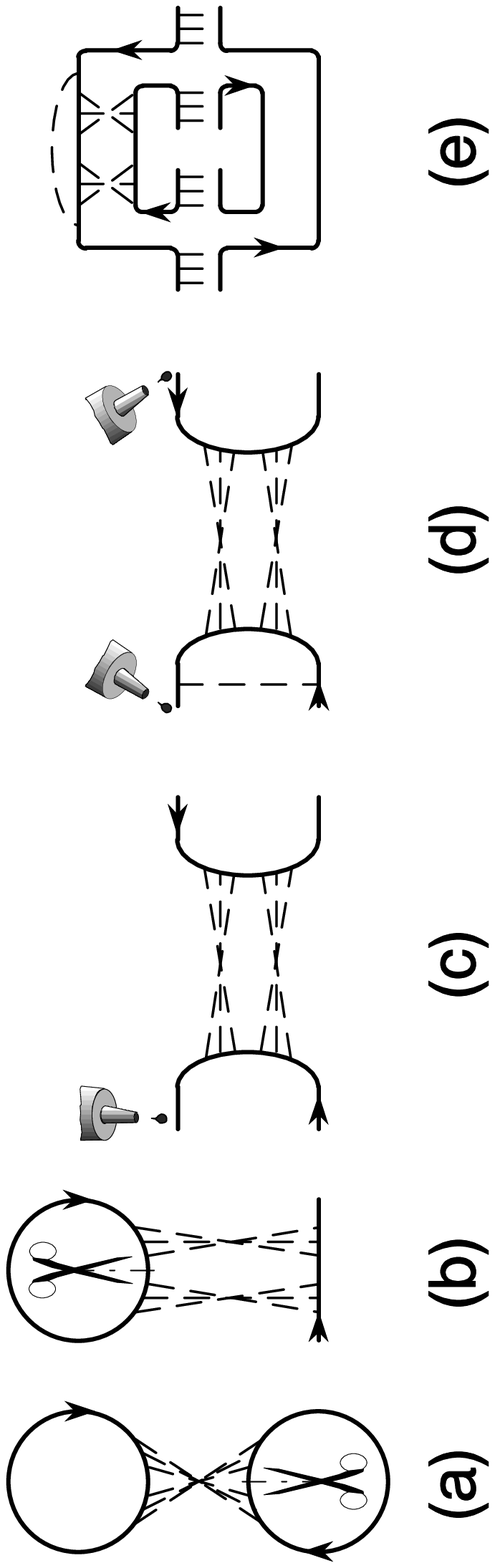,width=12cm}}
\caption{Particular example how a set of cooperons
 is produced.
     Functional differention with respect to G is indicated by cutting out
a line (pair
      of scissors) from (a) to (e). Pasting the result in the integral
equation,
     indicated in Fig.\ \protect{\ref{fi:r1}}, produces a four particle
quantitiy (and adds      external ladders).}
     \label{fi:r4}
\end{figure}

In the second example we start off with the diagram containing the maximally
crossed vertex, Figs.\ \ref{fi:r4}(a) or \ref{fi:r1}(c). By functional
differentiation, see Figs\ \ref{fi:r4}(b) and \ref{fi:r4}(c) and subsequent
integration in the Bethe-Salpheter equation (\ref{defgreen}) we obtain a
contribution which consists of two maximally crossed diagrams with a
dressing. This element is the input for Eq.(\ref{defdrieb}), see Fig.\
\ref{fi:r2}, and generates an important contribution to the whole set.

The set of diagrams thus generated by the Kadanoff-Baym approach was verified
using a computer program. We developed this program to generate all diagrams
with six or less scatterers automatically. After generating the set, the
program checks it against double counting.
Next, momenta are assigned to all propagators such that momentum
conservation is obeyed. It then determines whether a
particular diagram is of leading order in $(k\ell)^{-1}$; this is the case
if all propagators can have
a momentum approximately equal to $k$. The leading
diagrams are expressed in terms of the standard integrals defined in section
\ref{secshort}. The subsequent analytical calculation and summation of
diagrams was done by hand.  The program turns out to be especially useful in
determining the precise set of leading short-distance diagrams, but also the
long distance diagrams of section \ref{secset} were reproduced. It should be
mentioned that apart from the diagrams from the Kadanoff-Baym approach
we also found extra diagrams; in section \ref{secmbox} they are discussed
and it is shown that they cancel.

\section{Cancellation of short distance contributions}\label{secshort} In
this section we show that the strong divergence, that occurs if both Hikami
boxes act on internal diffusons, cancels. Since a diffuson decays rapidly if
its momentum becomes large, terms of the diffusons with few scatterers are
dominant when the momentum is large, which cause the divergence. The
cancellation is thus shown by considering in great detail the short distance
processes.

It turned out that complications arise for diagrams with  less than four
scatterers. The external diffusons may still contain an arbitrary number of
scatterers. For simplicity the calculation in this section is performed in
the bulk and the external diffusons are not attached to the diagrams. Since
the divergency is independent of the external momenta, the cancellation is
generally proven at zero external momentum. For simplicity we may then
consider an infinite system. As a result all factors can be expressed in
the d-dimensional momentum ${\bf q}=(Q,q_z)$. We first look at fixed value of
the $d$-dimensional internal momentum ${\bf q}$ and postpone the
integration.
We will show that under present conditions the integrand is zero
for all ${\bf q}$, so that there is no divergency after integration.

All diagrams can be factorized in products of the integrals \begin{equation}
I^{m,n}_{k,l} (q)=I^{k,l}_{m,n}(q)  \equiv \int \frac{d^3 {\bf
p}}{(2\pi)^3} G^k({\bf p})
G^{\ast l}({\bf p}) G^m({\bf p+q}) G^{\ast n}({\bf p+q}) \label{eqdefI}.
\end{equation}
In the calculation of the diffuson and the Hikami vertices in the previous
sections we expanded the integrals in $q \ell$,
Since we are after
contributions for $q\sim 1/\ell$, this expansion is not allowed.
We can still assume that $q \ll k $, as we do not need the
physics on length scales comparable to the wavelength, but comparable to
one mean free path only.
The integrals needed in the
calculation are given in Table \ref{tabeli}, where $A_i$ is an angular average
defined as \begin  {equation} A_i(q)=\frac{1}{4\pi} \int d\hat{{\bf p}}
\frac{1}{(1+\ell {\bf q} \cdot \hat{{\bf p}})^i}
\qquad(i=1,2,3).\end{equation} The
integral for $i=1$ yields the diffuson kernel $A_1= \arctan (q\ell)
/q\ell$. The internal diffuson now reads \begin{equation}{\cal
L}_{int}(q)=\frac{4\pi}{\ell}\frac{1}{1-A_1}.\label{eqladder2}
\end{equation}
We recover previous results by expansion in $(q\ell)$. This yield ${\cal
L}_{int}=12\pi/(\ell^3 q^2)$, agrees with the bulk solution of
(\ref{difeq}).
\subsection{Short distance properties of the long range diagrams}

First, look at the long range diagrams with few scatterings. By
writing out the dressing of the vertices one obtains the detailed structure
of the diagrams, given in Fig.~\ref{figregdet}. In the diagrams
Fig.~\ref{figreg}(a)i-(c)i, the incoming external diffusons are connected to
1 and 3. The diagrams in Fig.~\ref{figreg}(a)ii-(c)ii are obtained by
connecting external diffusons 1 and 2 to the incoming side. If
such a diagram contains only a few scatterers, it can be grouped into one of
the classes of the long range diagrams. A simple example is the upper left
diagram of Fig.~\ref{figregdet}, drawn there with six scatterers.
Interchanging the external diffusons numbered 2 and 3 clearly leads to
another topology. However, with only two scatterers present, the topology
does not change under this operation, and one must be careful not to
overcount this term. Bearing this important observation in mind, we sum
all long range diagrams for an arbitrary number of scatterers. The expression
for each diagram and its combinatorical factor is given in
Table~\ref{tablecom}. It is convenient to collect diagrams with an equal
number $m$ of scatterers that connect different propagators. Scatterers on
which
one given amplitude scatters twice are thus, momentarily, not counted as new
scatterers. Looking
at Fig.~\ref{figregdet}, $m$ equals the total number of scatterers minus the
number of scatterers indicated with curved lines. After this resummation we
find the important result that each class of ``long range'' terms vanishes
for $m$ not equal to 2. Denoting the terms of Fig.~\ref{figregdet}
respectively by $R_a$, $R_b$, and $R_c$ we find $R_a=R_b=R_c=0$, if
$m>2$. If
$m=2$, the classes cancel against each other, since we then obtain $R_a=R_b=-
R_c/2=\frac{\ell^4}{4k^4} A_1^2$. This implies  \begin{equation} R_a+R_b+R_c
= 0 .\end{equation} We have thus shown that the long range diagrams for an
infinite system cancel and thus cause no divergency after integration over
${\bf
q}$. It is essential that the degeneracy of the low order contributions be
counted properly.

Another, more standard way to verify the important cancellation is the
 following.
Recall that we look at a bulk situation with fixed internal momentum
${\bf q}$, while
the external momenta are put equal zero. Beyond the diffusion approximation
an internal diffuson is given by Eq.(\ref{eqladder2}).
We also need the Hikami-boxes beyond the diffusion approximation.  It holds
that \begin{equation} H_4({\bf q},0,-{\bf q},0)=I_{11}^{11}(q)
+\frac{4\pi}{\ell}\left\{I_{11}^{10}(q)\right\}^2
+\frac{4\pi}{\ell}\left\{I_{11}^{01}(q)\right\}^2. \end{equation}
Inserting the values from the table and doing similar but longer
calculations for two other vertices we find
\begin{mathletters}
\begin{eqnarray} H_4(0,{\bf q},0,-{\bf
q})&=&2H_4({\bf q},-{\bf q},0,0)=\frac{\ell^3}{8\pi k^2}
A_1(1-A_1), \\ H_6(0,0,{\bf q},0,0,-{\bf q})&=&
\frac{\ell^5}{16\pi k^4}A_1(1-A_1)(1-3A_1).
\end{eqnarray}
\end{mathletters}
Now the long range diagrams of Fig.~\ref{figreg} can be evaluated.
For the internal part one has at fixed ${\bf q}$
\begin{eqnarray} &2& H_4({\bf q},0,-{\bf q},0)^2{\cal L}_{int}^2(q)+
4 H_4({\bf q},-{\bf q},0,0)^2{\cal L}_{int}^2(q)+2 H_6({\bf q},0,0,-{\bf
q},0,0){\cal L}_{int}(q) \nonumber\\
&=&\frac{\ell^4}{4k^4}\{2A_1^2+A_1^2+2A_1(1-3A_1)\}
=\frac{\ell^4}{4k^4}(2A_1-3A_1^2)  \label{eqdiv2}.
\end{eqnarray}
It is essential that the denominators $1-A_1$ have disappeared from
this expression. As mentioned, it means that all high order terms cancel,
allowing a cancellation of the remainder by low order contributions.
{}From Table \ref{tablecom} it is seen
 that we have overcounted six types of low order terms.
The correction to be subtracted is
\begin{eqnarray} \label{divcor}
&&\frac{4\pi}{\ell}
\left\{4\frac{\ell^5}{32\pi k^4}(A_1+A_2+A_3)\right\} \nonumber \\
&+&\left(\frac{4\pi}{\ell}\right)^2
\left\{\left(\frac{\ell^3}{8\pi k^2}A_1\right)^2
+8\frac{-i\ell^4}{32\pi k^3}(2A_1+A_2)\frac{-i\ell^2}{8\pi k}A_1
+8\frac{i\ell^2}{8\pi k}\,\frac{i\ell^4}{32\pi k^3}(A_2+2A_3)
\right\}\nonumber\\
&+&\left(\frac{4\pi}{\ell}\right)^3 \left\{
8\frac{i\ell^2}{8\pi k}\,\frac{i\ell^2}{8\pi k}A_1
\frac{-\ell^3}{16\pi k^2}A_2+4\frac{-i\ell^2}{8\pi k}\,
\frac{i\ell^2}{8\pi k}\,
\frac{\ell^3}{8\pi k^2}A_3 \right\}
=\frac{\ell^4}{4k^4}(2A_1-3A_1^2).
\end{eqnarray}
It indeed exactly cancels expression (\ref{eqdiv2}), which arose as a
remainder of the long range terms.

{}From the results of present section one may be tempted to conclude that the
long range diagrams do not lead to divergencies if the correct degeneracies
of the low order terms is properly taken into account. Though this conclusion
is correct, it is too early to draw it, since we show in next parts that some
further complications arise.

\subsection{Extra short distance contributions}

\begin{figure} \psfig{file=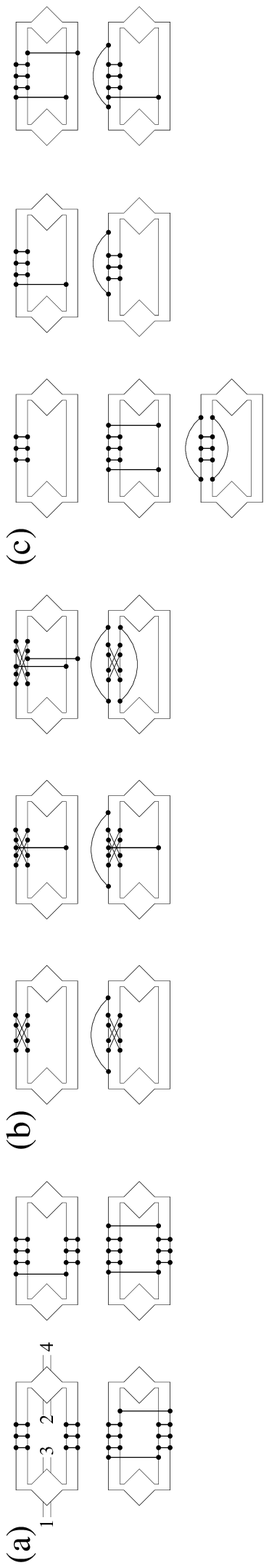,width=17.5cm} \caption{
The dressings of the shaded vertices of Fig.~\protect{\ref{figreg}} are
written out explicitely, this
shows the detailed structure of the long range contributions
to the conductance fluctuations.
The close parallel lines, only explicitly shown in the first diagram, are
the external diffusons.
Fig.~\protect{\ref{figreg}}(a)i corresponds to the case were 1 and 3 are
incoming diffusons; Fig.~\protect{\ref{figreg}}(a)ii corresponds to the case
were 1 and 2 are incoming diffusons. The vertical, diagonal or curved lines
linking two dots represent a common scatterering of two amplitudes. From top
to bottom the horizontal lines in a diagram are advanced, retarded, advanced
and, retarded propagators, respectively. The number of scatterers in the
internal and external diffusons is arbitrary. Topological equivalent diagrams
are not shown.} \label{figregdet}\end{figure}

\begin{figure} \psfig{file=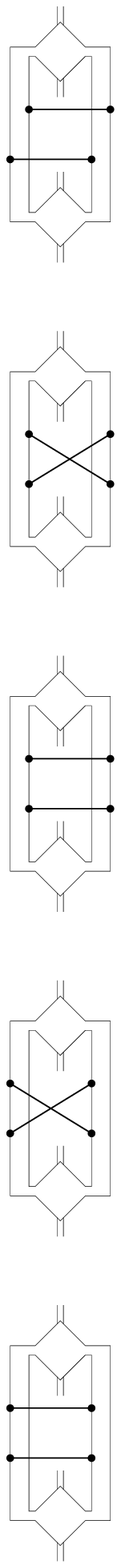,width=16cm} \caption{Leading
diagrams with topologies not yet contained in the long range diagrams. They
all contain two scatterers.} \label{figextra} \end{figure}

Unfortunately, we have not yet finished with the calculation as also
other terms are of leading order at short distances. First, diagrams with a
different topology also occur. There are five new diagrams, all with two
scatterers, see Fig.~\ref{figextra}. They can be looked upon as  diagrams
from the set of Fig.~\ref{figregdet}(a), but without any scatterers in the
ladders.

\begin{figure} \psfig{file=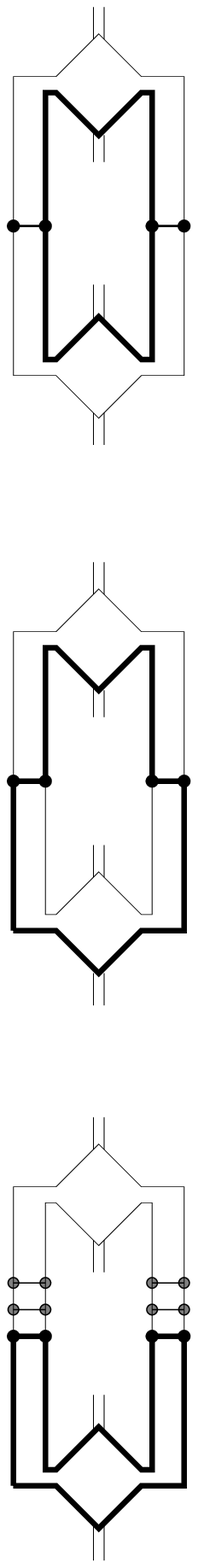,width=8cm} \caption{Different
choices for the loop momenta, all yielding leading contributions; thick and
thin lines depict different large momenta, both with length $k$. The first
diagram is leading for any number of scatterers. When only two scatterers
occur also the other two diagrams are leading.}\label{figmom} \end{figure}

Second, diagrams can have more than one configuration for a resonant
arrangement of the momenta, i.e. several configurations can be leading. The
most important contribution to the integral (\ref{eqdefI}) arises if the {\em
amplitude} loop momentum ${\bf p}$ is close to the wavevector $|{\bf p}|=k$.
With more loops present it is sometimes possible to find more than one choice
for the momenta of the amplitude propagators. This is illustrated in
Fig.~\ref{figmom}, where the diagram of Fig.~\ref{figregdet}(a)1 is repeated.
The expression for the shown diagram has contribution from three different
poles. The first arrangement of the momenta is dominant for any number of
scatterers, this one is part of the set of long range diagrams for any number
of scatterers. The two  configurations on the right give an extra leading
contribution. This only occurs with two scatterers present; with more
scatterers the two diagrams on the right become subleading, and only the left
diagram remains. Apart from this diagram, the same thing occurs for the
diagrams of Fig.~\ref{figregdet}(c)1 and Fig.~\ref{figregdet}(c)2 also with
two scatterers, and for the diagram Fig.~\ref{figregdet}(c)6 with three
scatterers.

Also the diagrams of Fig.~\ref{figextra} have more than one  pole: The four
left
diagrams of Fig.~\ref{figextra} have two resonant momenta configurations, the
right one has three. We sum all the contributions thus found (41 in total)
and denote them as
$S_n$ for $n$ scatterers. They are only non-zero for diagrams with two or
three scatterers, and read \begin{mathletters}
\begin{eqnarray} S_2&=&\left( \frac{4\pi}{\ell} \right)^2 \left[ 5
(I_{1,1}^{1,1})^2+ 8I_{1,1}^{1,1} I_{1,2}^{0,1}+12 (I_{1,2}^{0,1})^2
+8I_{1,1}^{0,1} I_{1,2}^{1,1} \right] =\frac{-\ell^4}{2k^4} A_1 A_2 , \\
S_3&=&8 \left( \frac{4\pi}{\ell} \right)^3 I_{1,1}^{0,1}I_{1,1}^{0,2}
I_{1,2}^{0,0} =\frac{\ell^4}{2k^4} A_1 A_2. \end{eqnarray} \end{mathletters}
One sees that these special contributions cancel.

\subsection{Interference vertices without partner exchange}\label{secmbox}
\begin{figure} \psfig{file=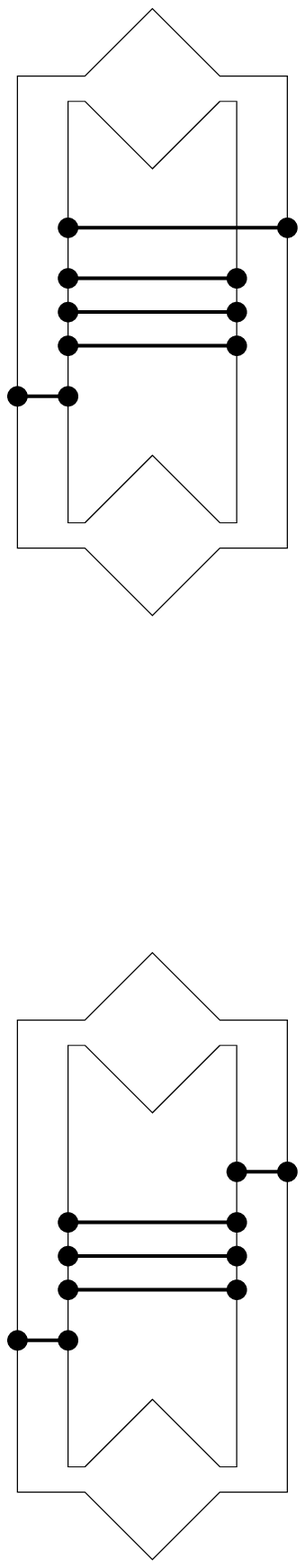,width=7cm} \caption{Leading diagrams
without amplitude exchange with one internal diffuson. The two drawn
diagrams cancel
against each other. All diagrams of this class cancel.}
\label{figmbox1} \end{figure}

\begin{figure}
\centerline{\psfig{file=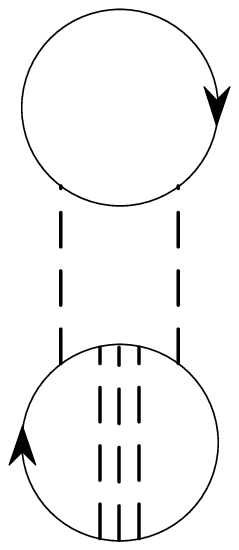,width=4cm}}
\caption{The generating functional that has to be included in the
Kadanof-Baym approach to generate the diagrams of
Fig.~\protect{\ref{figmbox1}}. } \label{figr5} \end{figure}

\begin{figure} \psfig{file=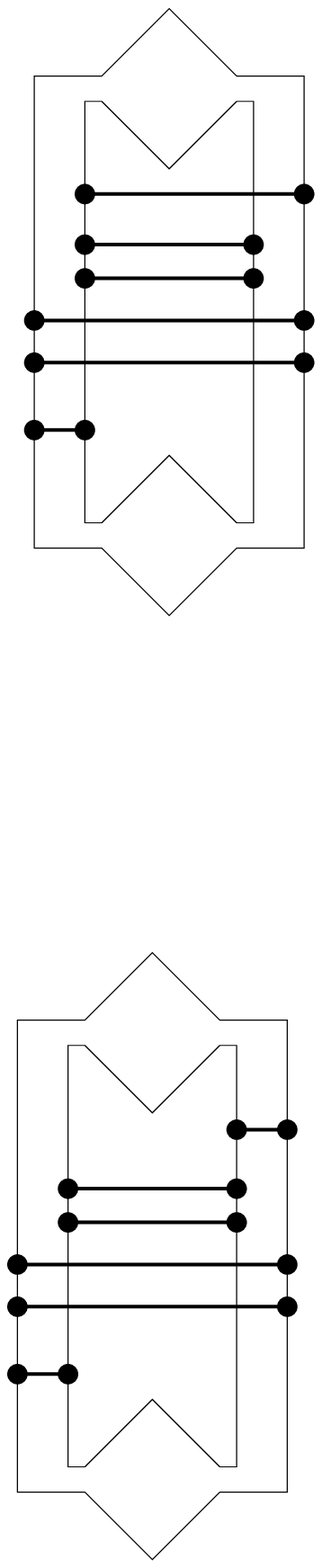,width=7cm} \caption{Leading diagrams
without amplitude exchange with two internal diffusons. The two drawn
diagrams cancel against each other. All diagrams of this class cancel.}
\label{figmbox2} \end{figure}

Apart from the diagrams generated by the Kadanoff-Baym approach, we found
another class of leading long range diagrams when we generated the diagrams
by computer. In these diagrams the two diffusons have two scatterers in
common, but no amplitude is exchanged. They either have one internal
diffuson, see Fig.~\ref{figmbox1} or two, see Fig.~\ref{figmbox2}. These
diagrams are actually of similar type as Fig.~\ref{figreg}. The diagrams with
one internal diffuson is analogous to Fig.~\ref{figreg}(c)i, while the ones
with two internal diffusons are analogous to Figs.~\ref{figreg}(a)i,(b)i.
The equivalents of Figs.~\ref{figreg}(a)ii,(b)ii and (c)ii also occur. The
important difference with the previous diagrams, though, is that at the
interference vertices
no amplitudes are exchanged. In this respect they differ qualitatively from
Hikami-boxes, where partner exchange does occur. Explicit calculation shows
that these terms cancel both at zero and non-zero external momentum, this
was also already noted by Kane, Serota and Lee\cite{kane}, see their
Fig.~5(c). These
classes of diagrams can thus be fully neglected. This cancellation is due to
time reversal invariance. Apart from these two classes, the Kadanoff-Baym
approach, using (\ref{deffunc}), generates all presented diagrams as we
checked with the computer program. The generating functional for the extra
diagrams of Fig.~\ref{figmbox1} is drawn in Fig.~\ref{figr5}.

In conclusion, the analysis of this section \ref{secshort} shows that when
the external momenta vanish, all leading terms cancel at fixed value of the
internal momentum. Upon integration over the loop momentum one still has zero
and, in particular, not a divergent contribution. For an infinite system the
theory is thus well-behaved at short distances. The divergence has canceled
in a careful study of the short-distance process; renormalization as
known from field theories was not needed.

\section{Application of conductance fluctuations to optical systems}
\label{seclong}
We have now seen that in infinite systems all divergent terms cancel. In
realistic systems, such as a slab, the same diagrams describe the relevant
physics. They have to be evaluated with appropriate diffuson propagators.
Generally they can be written as the bulk term with additional mirror terms.
For a
slab they were given in section \ref{secdif}. Knowing that the short distance
behavior is regular, we can selfconsistently consider all scattering diagrams
in the diffusion approximation. For the long range diagrams this was done
already in section \ref{secdifdiv}, where all long range contributions were
evaluated. The discussion of previous section has shown that the only new
effect comes from the subtraction of Eq. (\ref{divcor}). In the diffusion
limit for a quasi one-dimensional system at fixed transversal momentum $Q$ it
results in a contact term, labeled $F_d$.
\begin{equation} \label{eqFd}
F_d(M)=\int dz {\cal L}_{in}^2(z)\int_{-\infty}^\infty\frac{dq_z}{2\pi}
 \frac{\ell^4}{4k^4}{\cal L}_{out}^2(z)=\frac{\ell^4}{4k^4}\delta(0) \int
dz {\cal L}_{in}^2(z) {\cal L}_{out}^2(z).\end{equation}
This indeed cancels exactly the leading divergency that remained in
Eq.(\ref{Fabc}) for the sum of $F_a$, $F_b$ and $F_c$.
 The milder divergency present in the individual
terms $F_a$, $F_b$, $F_c$ in two and three dimensions
canceled already by summing them. Diagrammatically this can be seen as
follows: The diagrams responsible for the latter divergence
contain one internal diffuson (such in Fig.~\ref{figreg}(c)).
For these diagrams one does not have the complications
of the previous section; double counting corrections, different
resonant momentum configurations, and extra diagrams are absent. This
was confirmed by our computer generated diagrams.
The only contributions which remain in (\ref{eqsumf}) are thus the
double $z-$integrals. This finite remainder of our theory yields
the sought conductance fluctuations. We thus obtain, with again
$h_4=\ell^5/(48\pi k^2)$
\begin{eqnarray} F(Q,\kappa,\Omega)&=&F_{a}(M)+F_b(M)+F_c(M)+F_d(M)
\nonumber\\
&=&4 h_4^2\int \int \! {\rm d}z \,{\rm d}z' \,
{\cal L}_{int}(z,z';M) {\cal L}_{int}(z,z';M^\ast)\,
\left[ {\cal L}_{\rm in}'^2(z){\cal L}_{\rm out}'^2(z')+
{\cal L}_{\rm in}'(z) {\cal L}_{\rm out} '(z) {\cal L}_{\rm in}'(z')
{\cal L}_{\rm out} '(z') \right]\nonumber\\
&+&\frac{1}{2}h_4^2\int \int \! {\rm d}z \,{\rm d}z' \,
\left[{\cal L}_{int}^2(z,z';M)+{\cal L}_{int}^2(z,z';M^\ast)\right]
\frac{d^2}{dz^2}
\left[ {\cal L}_{\rm in}(z){\cal L}_{\rm out}(z)\right]
\frac{d^2}{dz'^2} \left[{\cal L}_{\rm in}(z') {\cal L}_{\rm out}(z')
\right]
\label{eqF},\end{eqnarray}
in which again $M^2=Q^2+\kappa^2+i\Omega$.
This equation is the central result of the present paper. The upper line of
(\ref{eqF}) corresponds to the diagrams of Fig.~\ref{figreg}(a), whereas the
lower line corresponds to the diagrams of Fig.~\ref{figreg}(b). Note that
only derivatives of external diffusons are present. As compared to the first
line, extra terms are present in the second one. According to
(\ref{difeq}) we have \begin{equation} \frac{d^2}{dz^2} \left[ {\cal
L}_{\rm in}(z){\cal L}_{\rm out}(z)\right]=2{\cal L}_{\rm in}'(z){\cal
L}_{\rm out}'(z)+2 \kappa^2 {\cal L}_{\rm in}(z){\cal L}_{\rm out}(z)
\label{eqLkappa} \end{equation}
The $\kappa-$terms are extra terms arising when absorption is present.
Finally, with Eq.(\ref{eqt2}) the value at vanishing
transversal momentum gives the variance of the conductance in one dimensions,
integration over the transversal momentum yields the correlation in two and
three dimensions.

Using the general result of (\ref{eqF}) and (\ref{eqt2})
various cases are considered by inserting the diffusons derived in section
\ref{secdif}. First consider the case of fully transmitting surfaces; if we
neglect absorption and frequency differences this gives
\begin{equation} F(Q)=\frac{3}{2} \frac{2+2Q^2
L^2-2\cosh 2QL+ QL\sinh 2QL}{Q^4 L^4 \sinh^2 QL} ,\end{equation}
 which decays for large $Q$ as $Q^{-3}$. In this case we recover
\begin{eqnarray} \langle
 T^2 \rangle_c & =&\frac{2}{15} \approx 0.133 ,\qquad \mbox{quasi 1d} \\
&=&\frac{3}{\pi^3} \zeta(3)\frac{W}{L}
 \approx 0.116 \frac{W}{L}
 , \qquad \mbox{quasi 2d} \\
&=&\frac{1}{2\pi}  \frac{W^2}{L^2}\approx 0.159\frac{W^2}{L^2} \qquad
\mbox{3d} ,\end{eqnarray} in which $\zeta$ is Riemann's zeta function.
These are well known results\cite{lee2}. We determine
also the frequency dependency of the correlation; this is important as it
determines the frequency range of the light needed to see the fluctuations.
Taking the frequency dependency into account we obtain \begin{eqnarray}
F(Q,\omega)&=& \frac{4\left( M^{\ast 2}-M^2 +M^2 M^\ast L \coth M^\ast L -
M^{\ast 2} ML \coth M L \right) }{L^4 M^2 M^{\ast 2} (M^2- M^{\ast 2})
}\nonumber \\  && + {\rm Re} \frac{2+2M^2 L^2-2\cosh 2ML+ ML\sinh 2ML}{2 M^4
L^4 \sinh^2 ML}. \label{fwq} \end{eqnarray} The correlation decays for large
frequency differences as $\Omega^{2-d/2}$, as was stated by Lee, Stone and
Fukuyama\cite{lee2}.

\begin{figure} \centerline{\psfig{file=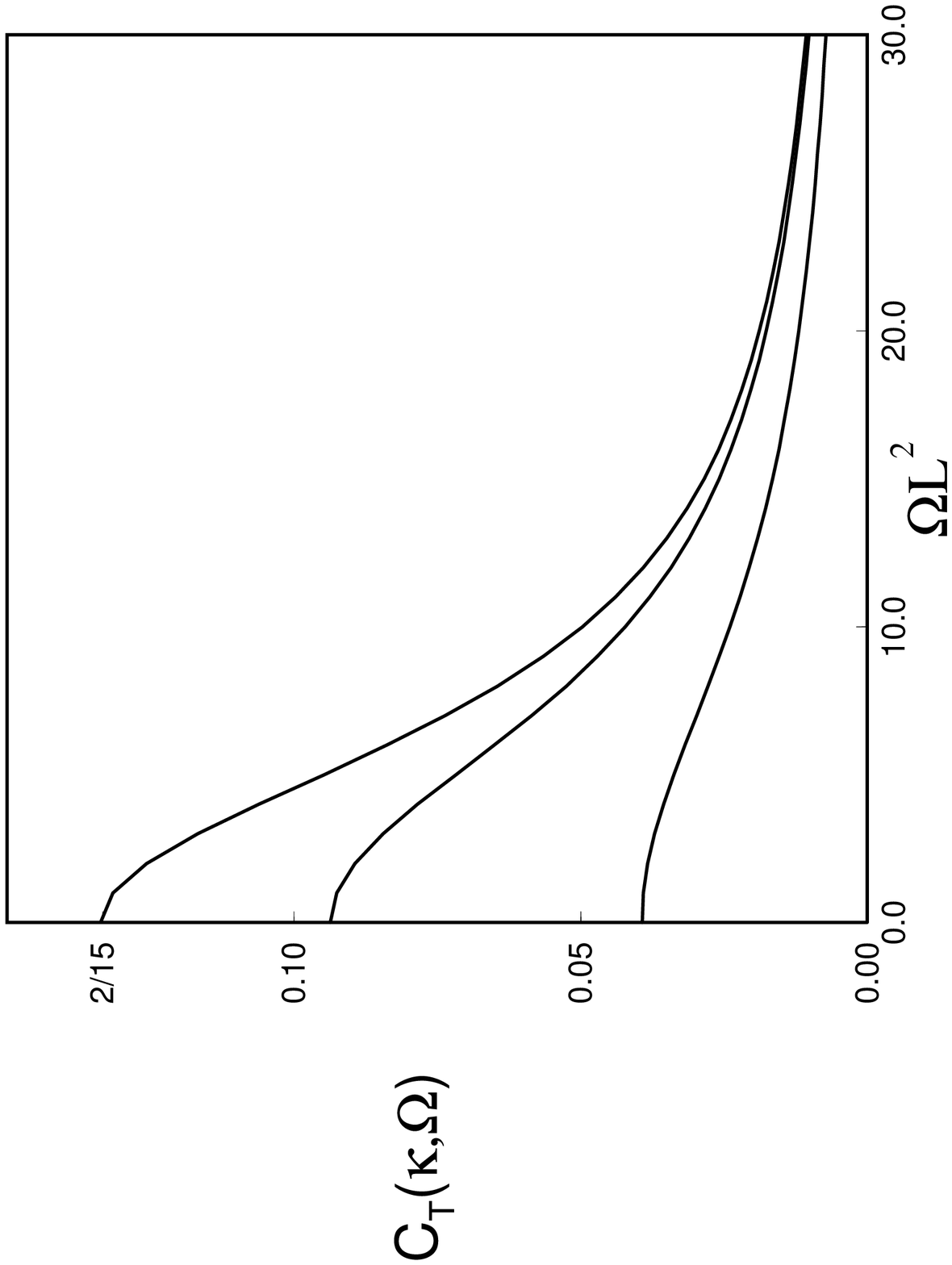,width=8cm}} \caption{The
correlation of the conductance as a function of the frequency for various
absorption strenghts in one
dimension without internal reflection. From upper to lower curve: no
absorption($\kappa=0$), $\kappa =1/L$ and, $\kappa=2/L$.} \label{figc3k1d}
\end{figure}

\begin{figure} \centerline{\psfig{file=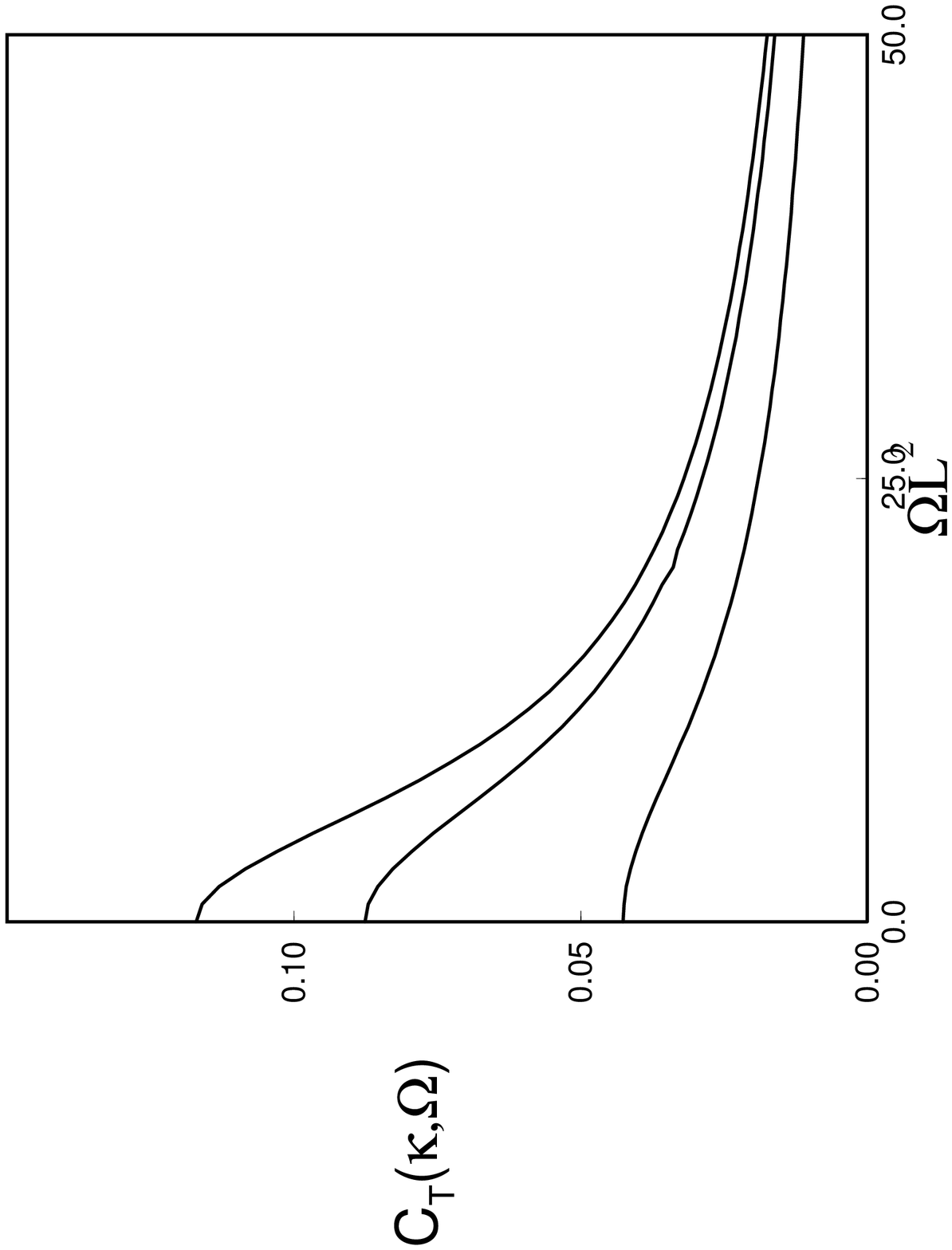,width=8cm}} \caption{The
correlation of the conductance as a function of the frequency for various
absorption strenghts in a two
dimensional slab without internal reflection. From upper to lower curve:
$\kappa=0$, $\kappa =1/L$ and, $\kappa=2/L$.} \label{figc3k2d} \end{figure}

\begin{figure} \centerline{\psfig{file=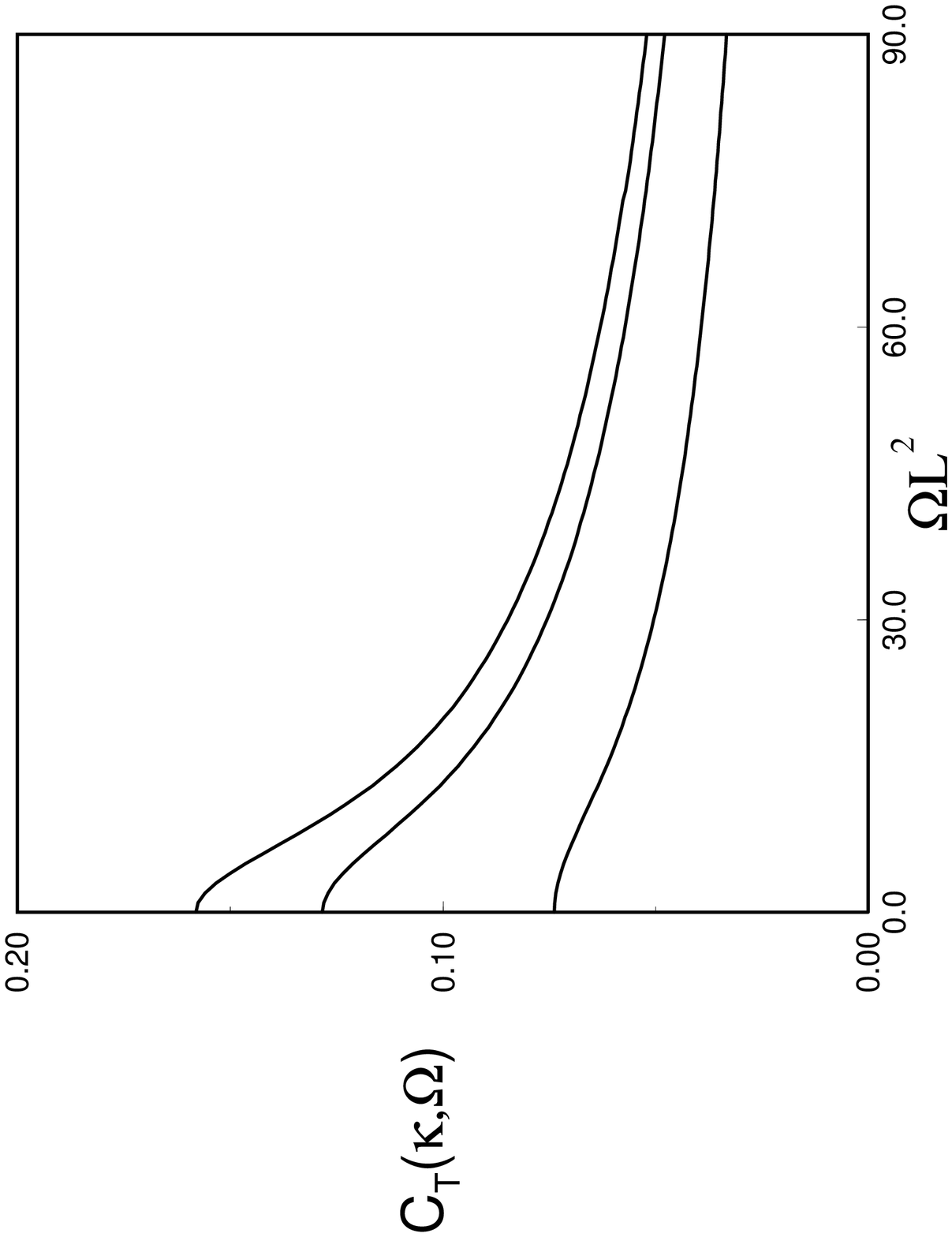,width=8cm}} \caption{The
conductance correlation function versus the frequency difference for various
absorption strenghts in a three
dimensional slab without internal reflection. From upper to lower curve:
$\kappa=0$, $\kappa =1/L$ and, $\kappa=2/L$.} \label{figc3k3d} \end{figure}

We were unable to perform the double integral over the position analytically
in presence of absorption. In Figs.~\ref{figc3k1d}, \ref{figc3k2d}, and
\ref{figc3k3d} we show the one, two, and three dimensional correlation
functions for various values of the absorption. It is seen that
especially the top of the correlation is reduced due to absorption.

\begin{figure} \centerline{\psfig{file=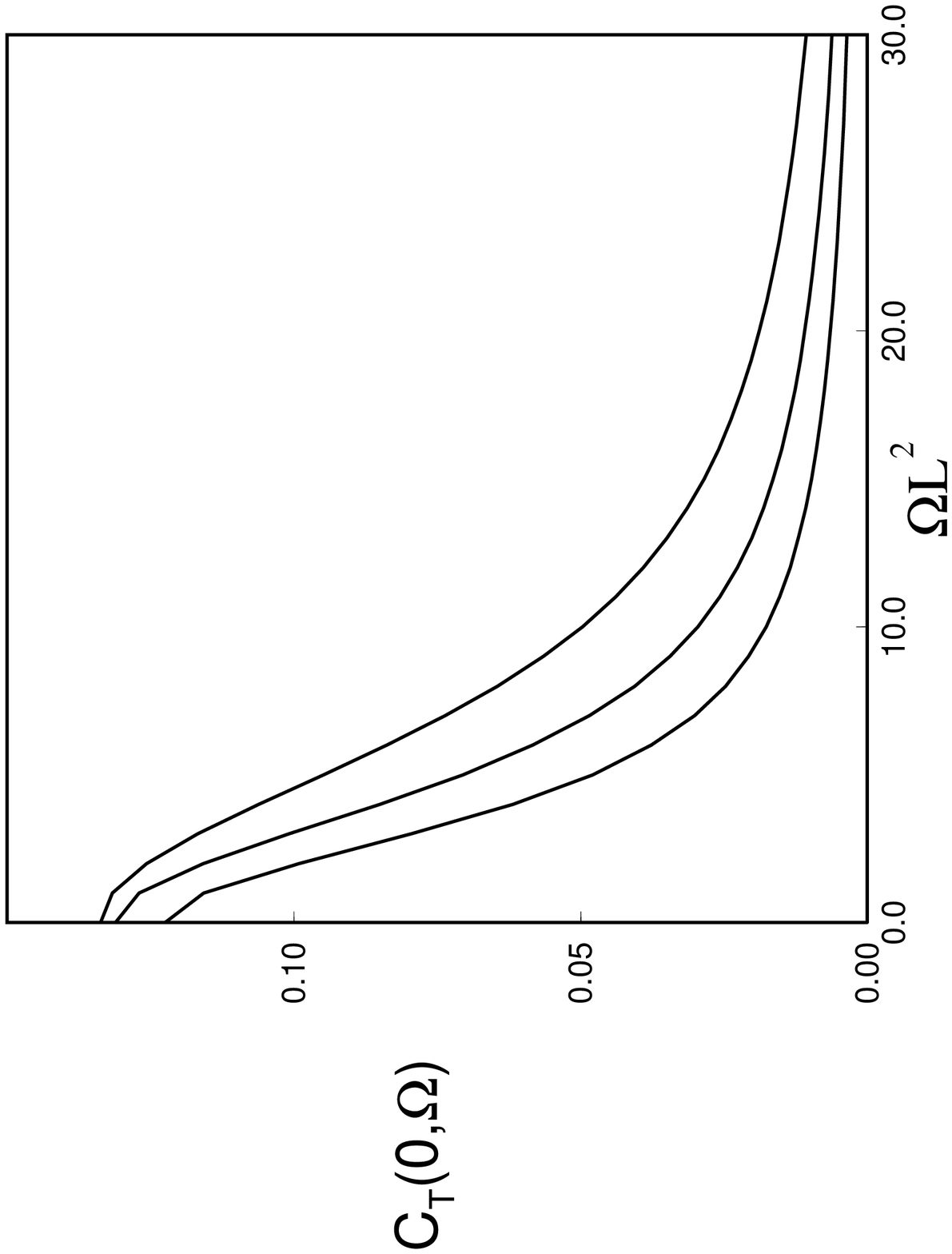,width=8cm}} \caption{The
influence
 of internal reflection on the one dimensional frequency correllation.
With $z_0=0, L/10, L/5 $ (upper, middle, lower curve); no absorption.
Also here the fluctuations
are reduced.} \label{figc3z01d} \end{figure}

\begin{figure}
\centerline{\psfig{file=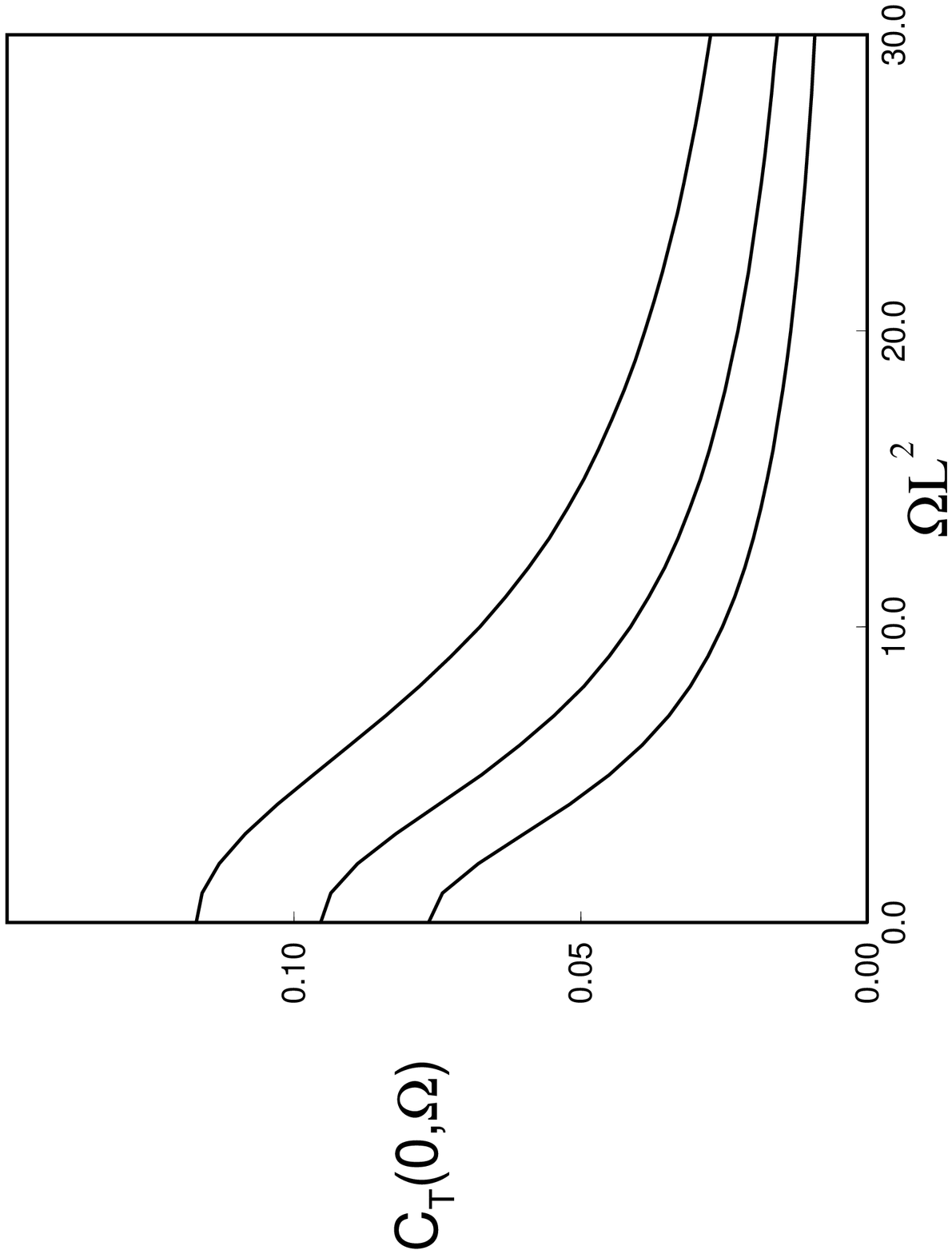,width=8cm}} \caption{Influence of
internal reflection on the frequency correllation in 2D; $z_0=0, L/10,
L/5 $ (upper, middle, lower curve); no absorption.} \label{figc3z02d}
\end{figure}

\begin{figure} \centerline{\psfig{file=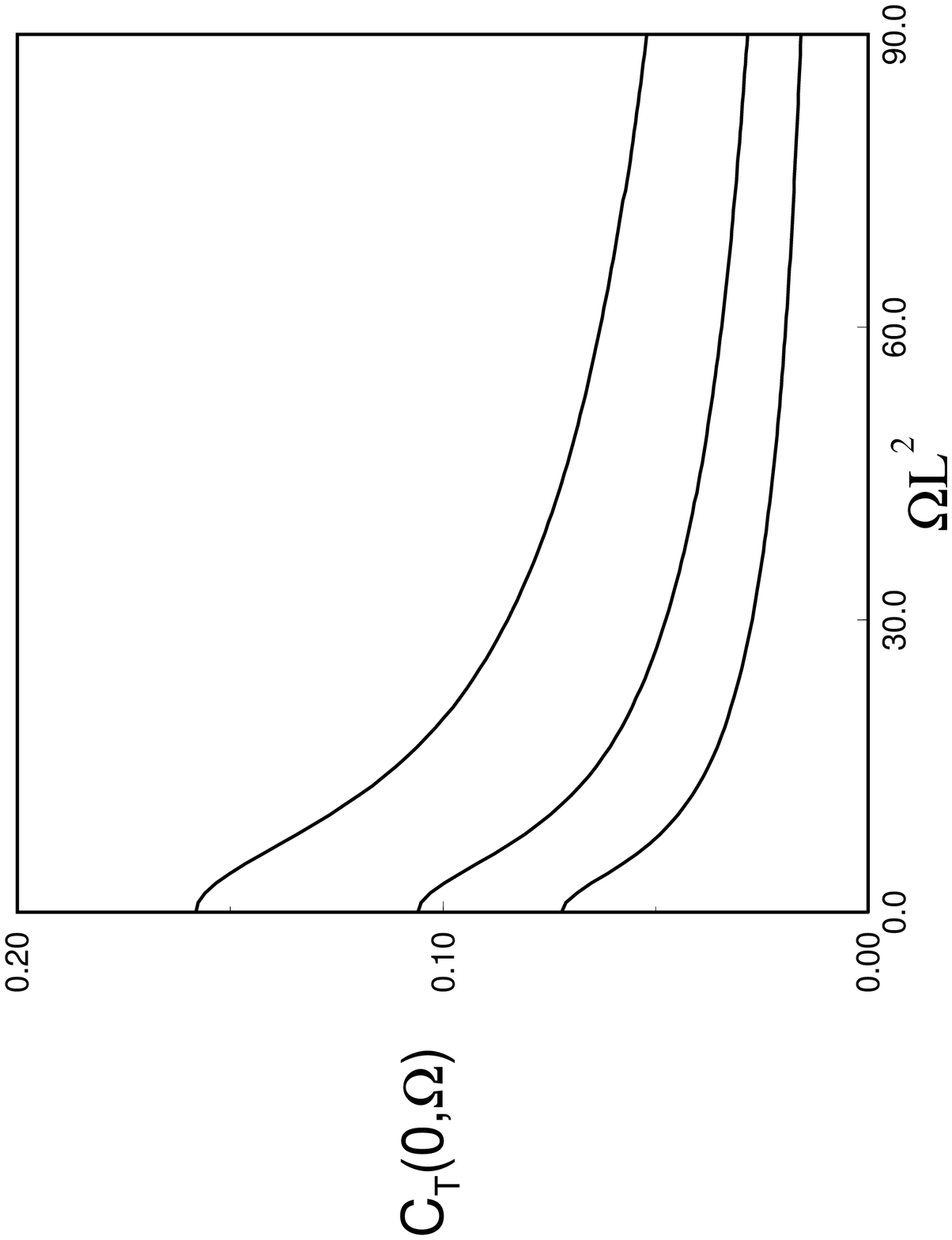,width=8cm}}
\caption{Influence of internal reflection on the frequency correllation in
3D;  $z_0=0, L/10, L/5 $ (upper, middle, lower curve); no absorption.}
\label{figc3z03d} \end{figure}

Next we applied our theory to the case of partial reflection at the surfaces
of the sample. We assume an index of refraction $m=\sqrt{\epsilon_0/
\epsilon_1} \neq 1$. For our purpose the internal reflections are coded in
only one parameter, the injection depth $z_0$, see (\ref{totfluxladin}),
(\ref{totfluxladout}), (\ref{Lint1}). For optically thick samples the
correlations are now determined by $z_0/L$. It is physically clear that the
internal reflections lead to a less steep diffuse intensity in the sample as
function of the depth. The fluctuations are proportional to the space
derivatives and are thus reduced. The results are presented in the
Figs.~\ref{figc3z01d}, \ref{figc3z02d} and \ref{figc3z03d} where the
correlation functions is plotted for various values of the ratio between
extrapolation length and sample thickness. One sees that already for small
ratios of $z_0/L$ the correlation is significantly lower than without surface
internal reflection.
It is seen
that neither the variance (the value at vanishing $\Omega$), nor the form of
the correlations are universal, they are sensitive to absorption and
internal reflections.

\section{Discussion and outlook}\label{secconlusion}

The universal conductance fluctuations (UCF) of mesoscopic electronic systems
have a direct counterpart in other mesoscopic systems with  multiply
scattered classical waves. The corresponding average normalized correlation
function is the $C_3$. It is known that a naive calculation
of this object in the Landauer approach is plagued with
short distance divergencies. In this work we have presented a detailed
diagrammatic approach to the calculation of the $C_3$ correlation function.
We first evaluated the leading long range diagrams.
As expected, each diagram, but also their sum, contains a short distance
divergency. Also a subleading divergency occurs, but for a slab geometry is
was found to cancel automatically. The study of the cancellation of the
leading divergency then was the main theme of the paper. Consistency of the
approach requires finding extra contributions that exactly cancel the already
determined divergency, and therefore serves as proof that the set of long
range diagrams exhaust all of them.

We developed a diagrammatic method that systematically generates all leading
scattering diagrams. This set was checked by computer. A large number of
diagrams had to be considered in detail. We summed these diagrams for an
infinite system at fixed value of the loop momentum. It was seen that also
beyond the diffusion approximation the action of Hikami boxes on the long
range propagators is to eliminate the long range terms, and  leave only
some low order contributions. Moreover, some low order terms of diagrams with
diffusons have a lower degeneracy than their higher order equivalents. Taking
this into account led exactly to a cancellation of all terms in an infinite
system. Next we discussed that some extra classes of leading diagrams occur,
but they all add up to zero. Thus in an infinite system all diagrams cancel,
so that, in particular, no short range divergency occurs. All short range
contributions could be coded in a contact term (\ref{eqFd}). It would be
interesting to investigate how it is derived in a non-linear sigma-model
formulation of the theory, possibly along the lines of Serota, Desposito and
Ma\cite{serota2}.

Subsequently the theory was applied to systems of finite size. Here the short
distance divergencies cancel as well, because the large scale geometry of the
system does not have influence on short range effects. The final result is
non-zero as it describes the correlation function in terms of derivatives of
external diffusion propagators of the geometry considered; such terms have no
meaning in an infinite system.

Our central result for the correlation function of the conductance is given
in (\ref{eqF}). It is obtained by adding the result of the long range
diagrams (\ref{Fabc})  and the contact term (\ref{eqFd}). When there is no
absorption, agreement is found with the result of Kane, Serota and
Lee\cite{kane}. All external diffusons are differentiated once. When
absorption is present, however, their approach is no longer valid. We found
that extra terms appear where some external diffusons are differentiated
twice or are proportional to $\kappa^2$, see (\ref{eqLkappa}).

We have applied the results to realistic optical systems. The frequency
dependent $C_3$ correlation function was calculated for the case where a
diffuse incoming beam is used and all outgoing intensity is collected. It was
seen that both absorption and internal reflections decrease the correlations
by a considerable amount. This is important for a quantitative analysis of
experimental data.

Electromagnetic measurements that involve the $C_3$ correlation have been
reported by Genack et al.\cite{garcia3}. These authors were able to describe
the data of their infrared experiments by adding the $C_1$, $C_2$ and $C_3$
contributions, but they incorrectly assumed that the $C_3$ is frequency
independent. The
experimental investigation of optical universal conductance fluctuations is
known to be very difficult. One problem is that if the incoming beam has to
be diffusive, it will have a low intensity.

We propose here a different way
to measure the same interference effect. Consider a laser beam coming in in a
given direction $a$ and  measure the frequency dependent total transmission.
Such can be done using an integrating sphere\cite{deboer}. Then repeat the
measurement for a very different incoming direction $c$. Each of these two
signals will exhibit the large $C_2$ correlation function\cite{deboer}.
However, when the directions $a$ and $c$ are not close to each other, the
$C_2$ will not contribute to the {\em cross correlate}. The cross correlation
of the total transmission
%$\sum_{b,b'} \langle T_{ab}\rangle \langle
%T_{ab'}\rangle C_3^{aabb'}= \langle T_{a}\rangle \langle T_{a}\rangle C_3
is much smaller than the autocorrelates; it just represents the typical
term of the UCF in (\ref{C3-G}). As this cross correlate is of relative
order $1/g^2$, it is of the same order of magnitude as the third cumulant of
the total transmission. A very precise measurement of that quantity was
carried out by de Boer, van Albada and Lagendijk, and reported in
collaboration with two of us\cite{t3prl}. It may therefore be expected that
it is just possible to measure $C_3$, and thus essentially the UCF, with
visible light.

\acknowledgements
M.v.R. thanks A. Genack, J. Hoogland, J. F. de Boer and R. Kop for
discussion.
Th.M.N. thanks P.A. Lee, B.L. Altshuler, I.V. Lerner and Shechao Feng for
discussions. R.V. kindly acknowledges helpful discussions with G.
Sch\"on on the theory presented in section \ref{secKB}. The research of
Th.M.N. was supported by the Royal Netherlands Academy of Arts and Sciences
(KNAW) and was also sponsored by NATO (grant nr. CGR 921399). The
research of R. V. was supported by the the Stichting voor Fundamenteel
Onderzoek der Materie (FOM).

\newpage


 

   

\newpage

\begin{table} \begin{tabular}{l|l} $I_{0,1}^{1,0} = \frac{\ell}{4\pi}
A_1$& $I_{0,2}^{1,0}= \frac{-i \ell^2}{8\pi k} A_2 $\\ $ I_{0,2}^{2,0}
 =\frac{\ell^3}{8\pi k^2} A_3 $& $ I_{1,1}^{0,1}= \frac{-i \ell^2}{8\pi k}
A_1 $\\ $ I_{1,1}^{0,2} = \frac{-\ell^3}{16\pi k^2} A_2 $& $ I_{1,1}^{1,1}
=\frac{\ell^3}{8\pi k^2} A_1$\\ $ I_{1,2}^{0,0} = \frac{-i \ell^2}{8\pi k} $&
$I_{1,2}^{0,1}= \frac{-\ell^3}{16\pi k^2} A_1 $\\ $ I_{1,2}^{1,0}
 =\frac{\ell^3}{16\pi k^2}[A_1+A_2] $& $I_{1,2}^{2,0}= \frac{i\ell^4}{32\pi
k^3} [A_2+2A_3]$\\$ I_{1,2}^{1,1} = \frac{-i \ell^4}{32\pi k^3} [2A_1+A_2]
$&$I_{1,2}^{2,1} = \frac{\ell^5}{32\pi k^4} [A_1+A_2+A_3]$ \end{tabular}
\caption{The short distance diagrams can factorized in these integrals,
defined in Eq.(\protect{\ref{eqdefI}});
$A_i$ is defined in
the text. \label{tabeli}} \end{table}

\begin{table}
\begin{tabular}{|l|l|cccc|c|}
Diagram &Expression&n=1 &$n=2$&$n=3$&$n\geq4$& $n-m$\\ \hline
a1 & $ \left[I_{1,1}^{1,1}\right]^2 \left[I_{0,1}^{1,0}\right]^{n-2} $
&0& 1=2-1& $4 $ & $2n-2 $ &0 \\
a2&$ I_{1,1}^{1,1}\left[I_{1,1}^{1,0}\right]^2
\left[I_{0,1}^{1,0}\right]^{n-3}$&0&0& $8$&$8n-16 $& 0\\
a3+a4 & $\left[I_{1,1}^{0,1}\right]^4 \left[I_{0,1}^{1,0}\right]^{n-4}$ &
0&0 & $0$ & $8n-24$&0 \\ \hline
b1 &$ I_{2,1}^{0,1} I_{1,2}^{1,0} \left[I_{0,1}^{1,0}\right]^{n-2}$ &
0&4&$8$&$4n-4 $&0 \\
b2 &$ I_{2,1}^{0,1}\left[I_{1,1}^{1,0}\right]^2
\left[I_{0,1}^{1,0}\right]^{n-3} $ &0& 0&8&$8n-16$&0 \\
b3 & $\left[I_{1,1}^{0,1}\right]^2 \left[I_{1,1}^{1,0}\right]^2
\left[I_{0,1}^{1,0}\right]^{n-4}$&0&0&0&$4n-12$ &0\\
b4 & $I_{2,1}^{0,0}
I_{1,2}^{1,0}I_{2,0}^{0,1}\left[I_{0,1}^{1,0}\right]^{n-3}$
&0&0&8&$8n-16$&1\\
b5 &$I_{2,1}^{0,0}I_{2,0}^{0,1}I_{1,1}^{1,0
}\left[I_{0,1}^{1,0}\right]^{n-4}$&0&0&0&$8n-24$&1\\
b6 &$I_{2,1}^{0,0}I_{2,0}^{0,2} I_{1,2}^{0,0}I_{0,2}^{1,0}
\left[I_{0,1}^{1,0}\right]^{n-4}$
&0& 0&0 &$4n-12$&2 \\ \hline
c1&$I_{1,2}^{2,1} \left[I_{0,1}^{1,0}\right]^{n-1} $&0=4-4& 4&4 &4 &0\\
c2&$I_{1,2}^{1,1}I_{1,1}^{0,1} \left[I_{0,1}^{1,0}\right]^{n-2} $ &0&
8=16-8&16&16 &0\\
c3&$I_{1,1}^{1,0}I_{1,1}^{0,1}I_{1,1}^{1,1}\left[I_{0,1}^{1,0}\right]^{n-3}$
&0&0 &8 &8&0\\
c4&$\left[I_{1,1}^{1,0}\right]^2 I_{2,1}^{1,0}
\left[I_{0,1}^{1,0}\right]^{n-3}$&0 &0 &8 &8&0\\
c5&$I_{2,1}^{0,0}I_{1,2}^{2,0} \left[I_{0,1}^{1,0}\right]^{n-2} $
&0&0=8-8&8 &8 &1\\
c6&$I_{2,1}^{0,0}I_{1,1}^{1,0}I_{2,0}^{1,1}
\left[I_{0,1}^{1,0}\right]^{n-3}$&0&0 &8=16-8 &16&1 \\
c7&
$I_{1,2}^{0,0}I_{2,1}^{0,0}I_{2,0}^{0,2}\left[I_{0,1}^{1,0}\right]^{n-3}$
&0&0&0=4-4 &4&2\\
\end{tabular}
\caption{Table used in the calculation of the diagrams presented on
Fig.~\protect{\ref{figregdet}} for zero external momenta. The expression for
each diagram is given in the second column. Its combinatorical prefactor in
the other columns for different number of scatterers. The last column counts
the number of scatterers that are not included as such in the resummmation.
There are six diagrams with degeneracy lower than expected: c1 for $n=1$,
a1,c2,c5 for $n=2$, and c6,c7 for $n=3$.} \label{tablecom} \end{table}

\end{document}